\documentclass[10pt,twocolumn,letterpaper]{article}

\usepackage{3dv}
\usepackage{times}
\usepackage{epsfig}
\usepackage{graphicx}
\usepackage{amsmath}
\usepackage{amssymb}

\usepackage{bm}
\usepackage{overpic}
\usepackage{wrapfig}
\usepackage{pgfplots}
\usepackage{pgf,tikz}

\usepackage{soul}
\usepackage{xcolor}

\usepackage[pagebackref=true,breaklinks=true,letterpaper=true,colorlinks,bookmarks=false]{hyperref}

\newcommand{\X}{\mathcal{X}}
\newcommand{\Y}{\mathcal{Y}}
\renewcommand{\v}{\mathbf{v}}
\newcommand{\R}{\mathcal{R}}

\newtheorem{lemma}{Lemma}
\newtheorem{property}{Property}

\threedvfinalcopy 

\def\threedvPaperID{145} 

\ifthreedvfinal\fi
\begin{document}

\title{Correspondence-Free Region Localization for Partial Shape Similarity\\via Hamiltonian Spectrum Alignment}

\author{
Arianna Rampini\\
Sapienza University of Rome\\
{\tt\small rampini@di.uniroma1.it}
\and
Irene Tallini\\
Sapienza University of Rome\\
{\tt\small tallini.1608460@studenti.uniroma1.it}
\and
Maks Ovsjanikov\\
LIX, \'Ecole polytechnique\\
{\tt\small maks@lix.polytechnique.fr}
\and
Alex M. Bronstein\\
Technion\\
{\tt\small bron@cs.technion.ac.il}
\and
Emanuele Rodol\`a\\
Sapienza University of Rome\\
{\tt\small rodola@di.uniroma1.it}
}

\maketitle

\begin{abstract}
We consider the problem of localizing relevant subsets of non-rigid geometric shapes given only a partial 3D query as the input. Such problems arise in several challenging tasks in 3D vision and graphics, including partial shape similarity, retrieval, and non-rigid correspondence. We phrase the problem as one of alignment between short sequences of eigenvalues of basic differential operators, which are constructed upon a scalar function defined on the 3D surfaces. Our method therefore seeks for a scalar function that entails this alignment. Differently from existing approaches, we do not require solving for a correspondence between the query and the target, therefore greatly simplifying the optimization process; our core technique is also descriptor-free, as it is driven by the geometry of the two objects as encoded in their operator spectra. We further show that our spectral alignment algorithm provides a remarkably simple alternative to the recent shape-from-spectrum reconstruction approaches. For both applications, we demonstrate improvement over the state-of-the-art either in terms of accuracy or computational cost.
\end{abstract}

\section{Introduction}
\vspace{-0.5ex}

Assessing similarity between non-rigid shapes is an active research topic in computer vision, pattern recognition and graphics \cite{biasotti2016recent}. At the heart of such methods lies the definition of shape {\em descriptors}, characterizing the shape either locally or globally (\eg, via the Bag-of-Words \cite{toldo2009bag} paradigm or via deep learning \cite{monti2017geometric}). Deformation-invariant shape descriptors often require careful tuning and hand-crafting, or sufficient training examples to enable learned-based methods.
While similarity by itself is usually expressed by a numerical score, the problem as a whole is also strictly related to (and often solved in tandem with) the complementary problem of shape {\em correspondence}. In this setting, local shape descriptors are used as `probe' quantities to employ in more sophisticated pipelines to infer a functional \cite{rodola2017partial} or point-to-point \cite{bronstein2009partial} correspondence.

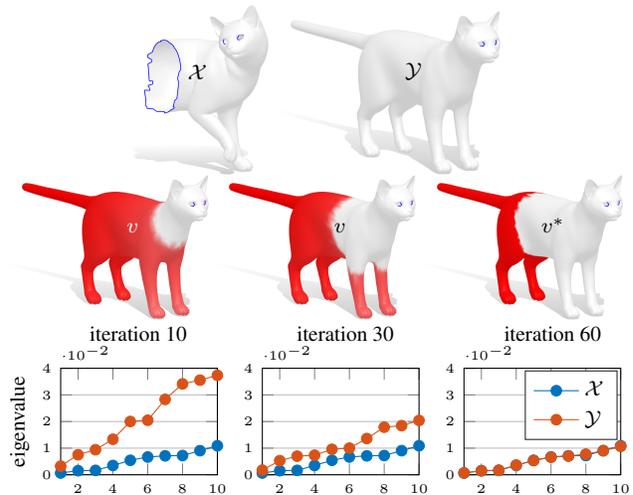
\begin{figure}[t!]
\centering
\begin{overpic}
[trim=0cm 0cm 0cm 0cm,clip,width=0.22\linewidth]{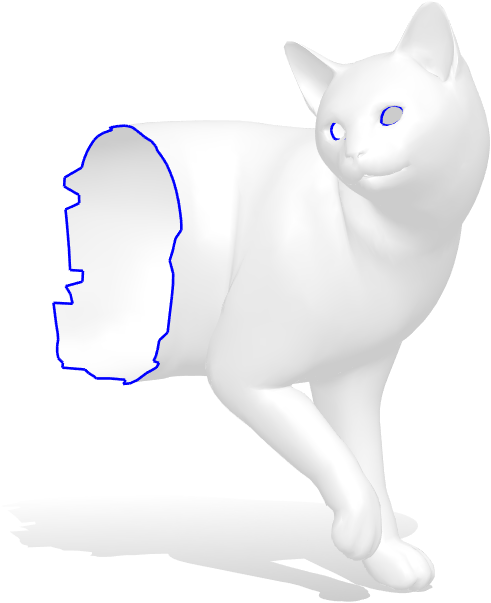}
\put(34,56){\footnotesize $\X$}
\end{overpic}
\begin{overpic}
[trim=0cm 0cm 0cm 0cm,clip,width=0.36\linewidth]{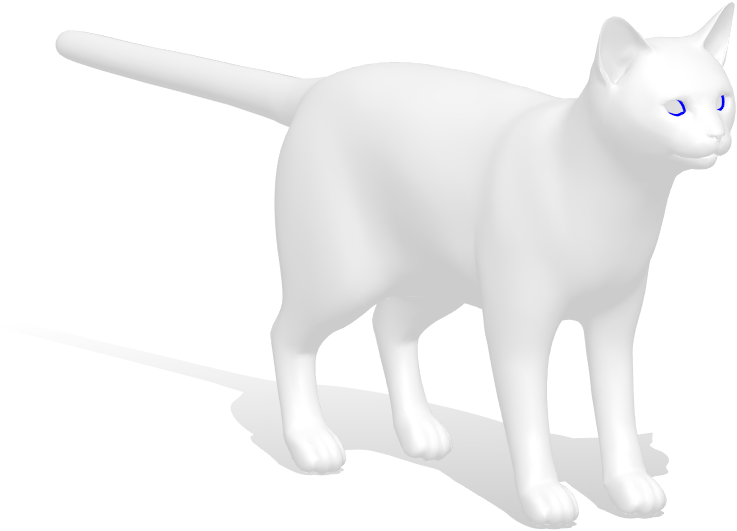}
\put(58,43){\footnotesize $\Y$}
\end{overpic}\\
  \begin{overpic}
[trim=0cm -3cm 0cm 0cm,clip,width=0.32\linewidth]{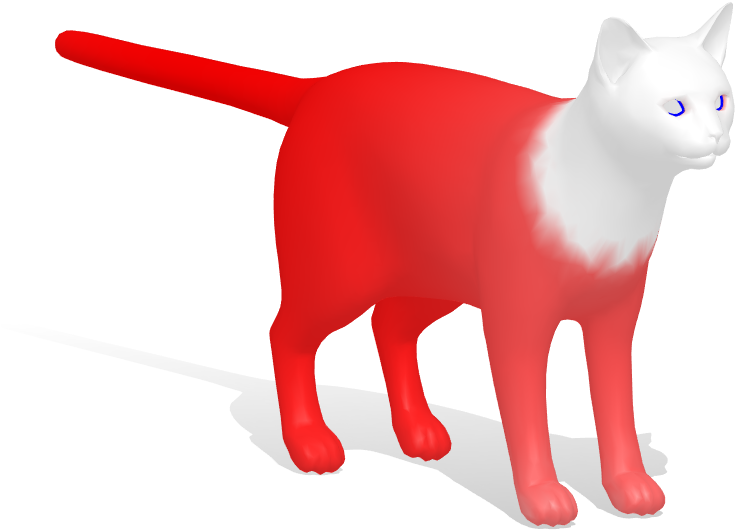}
\put(41,0.9){\footnotesize iteration 10}
\put(59.5,55){\footnotesize $\color{white} v$}
\end{overpic}
\begin{overpic}
[trim=0cm -3cm 0cm 0cm,clip,width=0.32\linewidth]{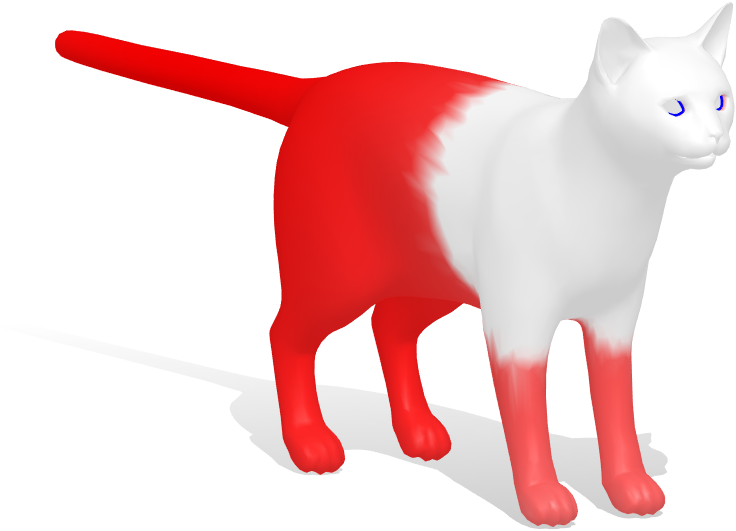}
\put(41,0.9){\footnotesize iteration 30}
\put(59.5,55){\footnotesize $\color{black} v$}
\end{overpic}
\begin{overpic}
[trim=0cm -3cm 0cm 0cm,clip,width=0.32\linewidth]{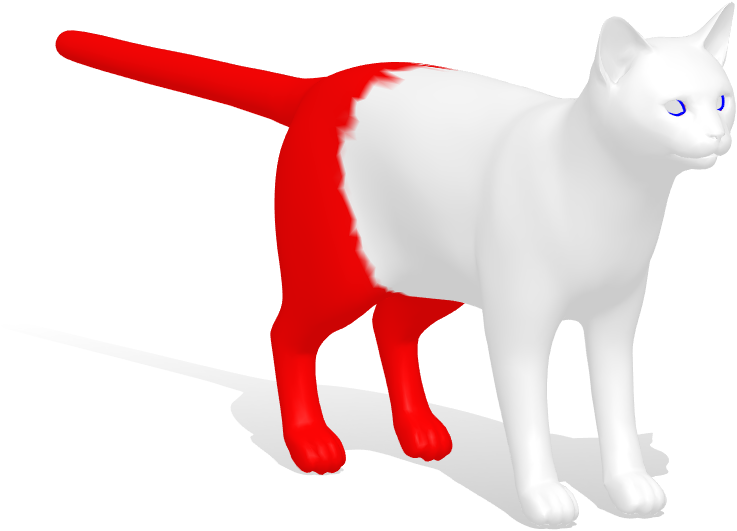}
\put(41,0.9){\footnotesize iteration 60}
\put(59.5,55){\footnotesize $\color{black} v^\ast$}
\end{overpic}
\setlength{\tabcolsep}{0pt}
  \begin{tabular}{ccc}
%
%
\definecolor{mycolor1}{rgb}{0.00000,0.44700,0.74100}%
\definecolor{mycolor2}{rgb}{0.85000,0.32500,0.09800}%
\begin{tikzpicture}

\begin{axis}[%
width=0.25\linewidth,
height=0.17\linewidth,
scale only axis,
xmin=1,
xmax=10,
ymin=0,
ymax=0.04,
ymajorgrids,
ylabel={\footnotesize eigenvalue},
ylabel style={font=\small, at={(axis description cs:0.21,0.5)},anchor=north},
every x tick label/.append style={font=\color{black}, font=\tiny},
every y tick label/.append style={font=\color{black}, font=\tiny},
axis background/.style={fill=white},
legend style={legend cell align=left, align=left, draw=white!15!black}
]
\addplot [color=mycolor1,solid,mark=*]
  table[row sep=crcr]{%
1	0.000660343424796039\\
2	0.00153453889959643\\
3	0.00165217212471269\\
4	0.00348054453326263\\
5	0.00542696818064847\\
6	0.00667355427297522\\
7	0.0071060820442721\\
8	0.00721619257486594\\
9	0.00904787303073995\\
10	0.010893666659463\\
};
\addlegendentry{target}

\addplot [color=mycolor2,solid,mark=*]
  table[row sep=crcr]{%
1	0.0032373381004831\\
2	0.0075218835430011\\
3	0.00940133628744455\\
4	0.0133331842678182\\
5	0.0200063421546712\\
6	0.0204746038422012\\
7	0.0283156388415833\\
8	0.0341368770254267\\
9	0.0355479553499407\\
10	0.0373609949591387\\
};
\addlegendentry{M}
\legend{}
\end{axis}
\end{tikzpicture}%
&
%
%
\definecolor{mycolor1}{rgb}{0.00000,0.44700,0.74100}%
\definecolor{mycolor2}{rgb}{0.85000,0.32500,0.09800}%
\begin{tikzpicture}

\begin{axis}[%
width=0.25\linewidth,
height=0.17\linewidth,
scale only axis,
xmin=1,
xmax=10,
ymin=0,
ymax=0.04,
ymajorgrids,
every x tick label/.append style={font=\color{black}, font=\tiny},
every y tick label/.append style={font=\color{black}, font=\tiny},
axis background/.style={fill=white},
legend style={legend cell align=left, align=left, draw=white!15!black}
]
\addplot [color=mycolor1,solid,mark=*]
  table[row sep=crcr]{%
1	0.000660343424796039\\
2	0.00153453889959643\\
3	0.00165217212471269\\
4	0.00348054453326263\\
5	0.00542696818064847\\
6	0.00667355427297522\\
7	0.0071060820442721\\
8	0.00721619257486594\\
9	0.00904787303073995\\
10	0.010893666659463\\
};
\addlegendentry{target}

\addplot [color=mycolor2,solid,mark=*]
  table[row sep=crcr]{%
1	0.00173983698673474\\
2	0.00530011048139656\\
3	0.00698267630367155\\
4	0.0073515516124889\\
5	0.00956455495804143\\
6	0.00999920032571389\\
7	0.0135917462379656\\
8	0.0178684529197071\\
9	0.0184444902741223\\
10	0.0203993305690897\\
};
\addlegendentry{M}
\legend{}
\end{axis}
\end{tikzpicture}%
&
%
%
\definecolor{mycolor1}{rgb}{0.00000,0.44700,0.74100}%
\definecolor{mycolor2}{rgb}{0.85000,0.32500,0.09800}%
\begin{tikzpicture}

\begin{axis}[%
width=0.25\linewidth,
height=0.17\linewidth,
scale only axis,
xmin=1,
xmax=10,
ymin=0,
ymax=0.04,
ymajorgrids,
every x tick label/.append style={font=\color{black}, font=\tiny},
every y tick label/.append style={font=\color{black}, font=\tiny},
axis background/.style={fill=white},
legend style={legend cell align=left, align=left, draw=white!15!black}
]
\addplot [color=mycolor1,solid,mark=*]
  table[row sep=crcr]{%
1	0.000660343424796039\\
2	0.00153453889959643\\
3	0.00165217212471269\\
4	0.00348054453326263\\
5	0.00542696818064847\\
6	0.00667355427297522\\
7	0.0071060820442721\\
8	0.00721619257486594\\
9	0.00904787303073995\\
10	0.010893666659463\\
};
\addlegendentry{\footnotesize $\X$}

\addplot [color=mycolor2,solid,mark=*]
  table[row sep=crcr]{%
1	0.000695273309730737\\
2	0.00147383672312884\\
3	0.00169172518381266\\
4	0.00363258952420331\\
5	0.00532935535674284\\
6	0.00659513458791494\\
7	0.00703372555893722\\
8	0.00774549635869803\\
9	0.00903262655045189\\
10	0.0106676683425775\\
};
\addlegendentry{\footnotesize $\Y$}

\end{axis}
\end{tikzpicture}%
  \end{tabular}
  \caption{\label{fig:teaser}
  Given a non-rigid partial 3D query $\X$ and a full shape $\Y$, we propose a way to locate $\X$ in $\Y$ without having to compute a correspondence $\pi:\X\to\Y$. We do so by looking for an indicator function $v$ on $\Y$ which, when used to construct a Laplacian-like matrix on $\Y$, its eigenvalues become the same as those of the classical Laplacian on $\X$. The optimal $v^\ast$ attaining the eigenvalue alignment localizes the sought region. Here we show the evolution of $v$ (middle row) and of the corresponding eigenvalues (bottom row) across our iterative algorithm until convergence.
  }
\end{figure}

Even more challenging is the setting of {\em partial} shape similarity, which is the focus of this paper. In this case, only some {\em regions} of the shapes are expected to be similar due to the presence of clutter or missing geometry.  Partial similarity arises in numerous practical tasks, ranging from object reassembly \cite{gregor2014towards,litany2016non} to protein docking \cite{shentu2008context} and deformable object-in-clutter detection \cite{cosmo2016matching} among others. 
In the partial setting, in addition to defining a similarity metric, one also has to identify the relevant similar regions of the given shapes, which is a difficult problem in itself (see Figure~\ref{fig:teaser}). As a result, while global similarity has been extensively studied, only few methods exist to address the (especially non-rigid) partial 3D similarity problem.
%

A seemingly unrelated problem concerns the recovery of the geometry of an unknown shape (\ie, whose point coordinates in space are not given) from a minimal amount of input data. Notably, it has been recently shown \cite{isosp} that, in some cases, one can reconstruct the geometry of a shape even when an incomplete sequence of its Laplacian eigenvalues (its spectrum) are given as an input -- a problem widely known in mathematical physics as ``hearing the shape of the drum''~\cite{kac}. In vision and graphics, this problem finds applications in non-isometric shape matching  as well as style transfer \cite{isosp} and acoustics design \cite{bharaj2015computational}.

In this paper we show that partial shape similarity and shape-from-spectrum recovery, while apparently very distinct, can be phrased as instances of the same mathematical problem. We formulate this problem by exploiting the notion of the Hamiltonian operator on manifolds, and reduce it to an unconstrained continuous optimization problem over the vector space of real-valued functions defined over the surfaces; see Figure~\ref{fig:teaser} for an illustration of our methodology on a real example.


\subsection{Related work}\label{sec:rel}
Partial similarity of 3D shapes has long been researched by the vision and geometry processing communities, with the vast majority tackling the {\em rigid} setting. 
 Since the focus of this work is on {\em non-rigid} structures, we discuss methods addressing deformable objects here, and refer the interested reader to the recent surveys \cite{sipiran2013shrec,pratikakis2016partial,phamshrec} for the former case.

\vspace{1ex}\noindent\textbf{Partial shape retrieval.}
%
At a high level our task is related to shape retrieval, where the goal is to produce a {\em ranking} of most similar shapes according to a similarity score.
Even in this simpler task, surprisingly few methods have attempted to address the  problem.  We mention here the recent SHREC'17 challenge \cite{shrec17}, highlighting how deep learning-based methods may work well for producing rankings, provided that enough training examples are available (the organizers provided around two thousands).

Our task is more challenging, as we aim at identifying the regions that two non-rigid objects have in common.

\vspace{1ex}\noindent\textbf{Partial shape correspondence.}
When a map $\pi:\X\to\Y$ between shapes $\X$ and $\Y$ is available, one may simply look at the image $\pi(\X)\subset\Y$ in order to determine the corresponding region (for simplicity, here we assume that $\X$ is a deformed region of $\Y$ as in Figure~\ref{fig:teaser}). Thus, any partial shape matching method implicitly solves for the common region as a side product, at the cost of solving a computationally heavy correspondence problem. 


To date, solving a full correspondence problem is therefore the dominant approach. Bronstein~\etal \cite{bronstein2009partial} proposed to explicitly solve for a point-to-point map $\pi$ and a membership function $\mu$ (or ``fuzzy part'', measuring the degree of inclusion of each point into the subset), optimizing for $\pi$ and $\mu$ in an alternating fashion. Optimization for $\pi$ was posed as a NP-hard quadratic assignment problem \cite{burkard1998quadratic}, while solving for $\mu$ required solving a non-convex quadratic problem due to the presence of region regularizers.
More recently, Rodol\`a~\etal \cite{rodola2017partial} followed a similar approach where the correspondence step was rephrased using the functional map representation \cite{ovsjanikov2012functional}. A membership function is still optimized for, with the same regularizers as those in \cite{bronstein2009partial}. The framework was extended in \cite{litany2016non} to address the case with $n>2$ regions (``non-rigid puzzle''), leading to a multi-label segmentation problem where $n$ functional maps are optimized simultaneously. The presence of clutter was investigated in \cite{cosmo2016matching} using the same formulation of \cite{rodola2017partial}, but where the local descriptors (which are used in all the methods above to drive the correspondence) are learned via deep metric learning on a representative dataset.
Finally, Brunton~\etal \cite{brunton2014low} proposed an iterative approach where a partial map is recovered by an isometric growing procedure starting from sparse pre-matched features over the surfaces.

In contrast to these methods, our approach does not require solving for dense correspondences, thus essentially eliminating the need for local descriptors, and does not make use of complex regularizers. Instead, we phrase the optimization problem in terms of solving for a real-valued (indicator) function aimed at identifying similar regions, and devise a correspondence-free metric that allows us to measure and optimize for the similarity of such functions directly.

A related approach was previously introduced by Pokrass~\etal \cite{pokrass2013partial} for identifying region-wise similarity. However, unlike our method, that approach  relies on the comparison of \emph{descriptor} statistics over the surface, thus strongly depending on the specific choice of descriptors.
%

\vspace{1ex}\noindent\textbf{Isospectralization.}
From an algebraic standpoint, ours may be seen as an ``inverse eigenvalue problem''~\cite{chu2005inverse}; despite the name, problems in this domain typically require knowledge of the eigenvectors of the sought operator, while our method does not. Perhaps, the most closely related to ours is the recent work on \emph{isospectralization} \cite{isosp} where the authors  deform the shapes with the goal of aligning their Laplacian eigenvalues. We also use a similar observation that in practice these eigenvalues provide a powerful characterization of the shape's geometry. However, we focus on the partial setting and propose to solve for a characteristic function of a region, thus avoiding any explicit shape deformation.

\subsection{Contribution}
%
The key contribution of this paper is to show that for many practical problems, it is possible to side-step the need for a correspondence by looking at 
 the eigenvalues of certain differential operators constructed on each surface. 
 This leads to tangible consequences that we leverage in two challenging applications. Specifically,
\begin{itemize}
\item We introduce a new method for detecting similar regions among deformable shapes; for the first time, the optimization is completely {\em correspondence-free} and, at its core, descriptor-free;
\item Our optimization problem uses a single objective, thus avoiding possibly complex regularizers, and can be solved {\em efficiently} with off-the-shelf differentiable programming libraries; 
\item We provide a remarkably simple alternative to recent {\em isospectralization} techniques \cite{isosp}, while at the same time yielding qualitatively better results at a fraction of the time cost.
\end{itemize}

\section{Background}
\vspace{-0.5ex}
%
%
%
In this paper we model shapes as 2-dimensional Riemannian manifolds $\X$ (possibly with boundary $\partial\X$) equipped with the area element $\mathrm{d}x$. We denote by $\mathrm{int}(\X)$ the interior of $\X$, namely the set of points $\X\setminus\partial\X$. 

\vspace{1ex}\noindent\textbf{Laplacian.}
The positive semi-definite Laplace-Beltrami operator $\Delta_\X$ (or manifold Laplacian) generalizes the basic differential operator from Euclidean analysis to Riemannian manifolds. It admits a spectral decomposition
\begin{align}
\Delta_\X \phi_i(x) &= \mu_i \phi_i(x)  &x\in\mathrm{int}(\X) \label{eq:eig} \\
\phi_i(x) &=0  &x\in\partial\X\,,\label{eq:d}
\end{align}
with homogeneous Dirichlet boundary conditions
\eqref{eq:d}, where $\{ 0<\mu_1 \leq \mu_2 \leq \dots \}$ are the eigenvalues (each with finite multiplicity) and $\phi_i$ are the associated eigenfunctions. For the remainder of the paper, we always assume eigenvalues to be increasingly ordered.

The {\em spectrum} of $\Delta_\X$ is the set of all eigenvalues, forming a discrete subset of $\mathbb{R}_+$. It encodes geometric and topological properties of $\X$; for instance, it has been used as a basic global shape descriptor for 3D shape retrieval \cite{reuter2006laplace}.

A direct consequence of Eq.~\eqref{eq:eig} is that each eigenvalue $\mu_i$ is associated to an eigenfunction $\phi_i$ by the equality\footnote{In this equality we use the intrinsic gradient operator $\nabla_\X$ acting on scalar functions $f:\X\to\mathbb{R}$, which defines the direction (tangent to the surface) of the steepest change of $f$ around each point $x\in\X$.}
\begin{align}\label{eq:de}
\mu_i = \int_\X \| \nabla_\X \phi_i (x) \|^2_2 \,\mathrm{d}x 
\end{align}
after normalization so that $\int_\X \phi_i^2(x) \mathrm{d}x =1$. Equation~\eqref{eq:de} makes the {\em global} nature of the eigenvalues clear, since they are computed as integrals over the entire surface.

\vspace{1ex}\noindent\textbf{Hamiltonian operator.} 
Given a scalar {\em potential} function $v:\X\to\mathbb{R}_+$, one can define a Hamiltonian  $H_\X = \Delta_\X + v$, operating on scalar functions $f$ as follows:
\begin{align}
H_\X f = \Delta_\X f + v f\,,
\end{align}
where the product $vf$ is taken point-wise. Trivially, for the null potential where $v(x)=0 ~~\forall x\in\X$, the Hamiltonian $H_\X$ simplifies to $\Delta_\X$.
The operator $H_\X$ also admits a spectral decomposition:
\begin{align}
H_\X \psi_i(x) = \lambda_i \psi_i(x)\,,
\end{align}
with boundary conditions when necessary. Hamiltonian eigenfunctions have been used in shape analysis applications in \cite{choukroun2018hamiltonian,melzi2018localized,rodola2019functional}.

\vspace{1ex}
%
%
%
%
%
\setlength{\columnsep}{7pt}
\setlength{\intextsep}{1pt}
\begin{wrapfigure}[4]{r}{0.23\linewidth}
\vspace{-0.6cm}
\begin{center}
\begin{overpic}
[trim=0cm 0cm 0cm 0cm,clip,width=1.0\linewidth]{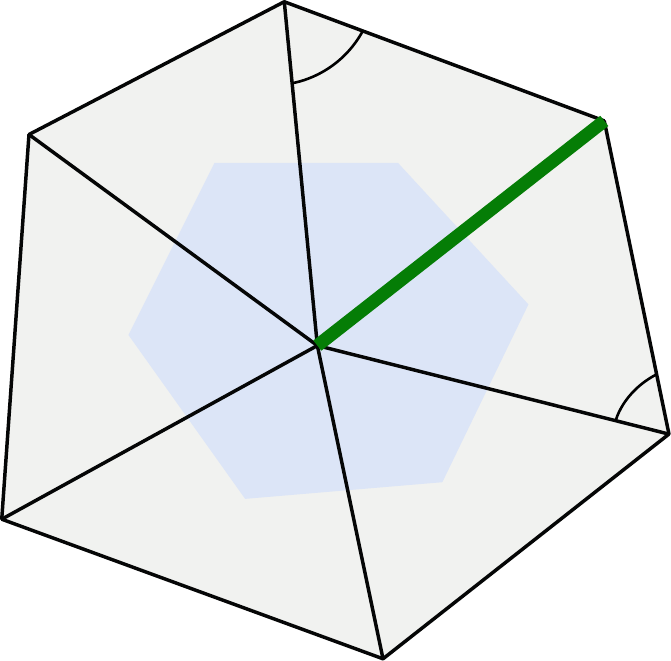}
\put(29,96){\footnotesize $h$}
\put(34,45){\footnotesize $i$}
\put(100,28){\footnotesize $k$}
\put(93,77){\footnotesize $j$}
\put(48,80){\footnotesize $\alpha_{ij}$}
\put(72,45){\footnotesize $\beta_{ij}$}
\end{overpic}
\end{center}
\end{wrapfigure}
\noindent\textbf{Discretization.} In the discrete setting, $\X$ is approximated by a triangle mesh with $n$ vertices $v_i\in V_\mathrm{int} \cup V_\mathrm{bdr}$, and where each edge $e_{ij}\in E_\mathrm{int} \cup E_\mathrm{bdr}$ belongs to at most two triangle faces $F_{ijk}$ and $F_{jih}$ (see inset for notation). We use the subscripts $\mathrm{int}$ and $\mathrm{bdr}$ on vertices and edges to denote interior and boundary, respectively.

The Laplacian is defined in terms of two $n\times n$ matrices $\mathbf{W}$ and $\mathbf{A}$, 
 where $\mathbf{A}$ is a diagonal matrix of local area elements $a_i$ 
 and $\mathbf{W}$ is a symmetric matrix of edge-wise weights (also known as cotangent formula, see \eg \cite{meyer2003discrete}):
\begin{eqnarray}
\label{eq:cotan}
w_{ij} \hspace{-0.3cm} &=& \hspace{-0.4cm} \left\{ 
		\begin{array}{lc}
{\scriptstyle -(\cot \alpha _{ij} + \cot \beta _{ij})/2} & {\scriptstyle e_{ij} \in E_\mathrm{int};}  \\
{\scriptstyle 0} & {\scriptstyle (i\neq j) \wedge (i \in V_\mathrm{bdr} \vee j \in V_\mathrm{bdr})}; \\
{\scriptstyle \sum_{k\neq i} (\cot \alpha _{ik} + \cot \beta _{ik})/2} & {\scriptstyle (i = j) \wedge v_i\in V_\mathrm{int};} \\
{\scriptstyle 1} & {\scriptstyle (i = j) \wedge v_i\in V_\mathrm{bdr};}
		\end{array}
\right.\\
a_i  \hspace{-0.3cm} &=& \hspace{-0.4cm} \left\{ 
		\begin{array}{lc}
			\frac{1}{3}\sum_{jk:ijk\in F} A_{ijk} &   v_i \in V_\mathrm{int};  \\
			0 &   v_i \in V_\mathrm{bdr};
		\end{array}
\right.
\end{eqnarray}
where $A_{ijk}$ is the area of triangle $F_{ijk}$. In the formulas above, the special treatment of boundary edges and vertices imposes the Dirichlet boundary conditions of Eq.~\eqref{eq:d}. 

%
A generalized eigenproblem $\mathbf{W}\bm{\Phi} =  \mathbf{A}\bm{\Phi}\mathrm{diag}(\bm{\mu})$ is solved for computing the Laplacian eigenvalues and eigenvectors. In the case of the Hamiltonian, it takes the form
\begin{align}\label{eq:eigs}
(\mathbf{W}+\mathbf{A}\mathrm{diag}(\v))\bm{\Psi} =  \mathbf{A}\bm{\Psi}\mathrm{diag}(\bm{\lambda})\,.
\end{align}
Above, $\mathbf{v}$ is a $n$-dimensional vector containing the values of the potential $v$ at each vertex, $\bm{\Phi}$ and $\bm{\Psi}$ are matrices containing the first $k$ eigenvectors as their columns, and $\bm{\lambda},\bm{\mu}$ are the corresponding $k$-dimensional vectors of eigenvalues. 

For brevity, we will write $\bm{\lambda}(\bm{\Delta}_\X+\mathrm{diag}(\v))$ to denote the eigenvalues appearing in Eq.~\eqref{eq:eigs}.
\section{Summary of our reasoning}
\vspace{-0.5ex}

\begin{figure}[t]
  \centering
  \setlength{\tabcolsep}{0pt}
  \begin{tabular}{ccc}
%
%
\begin{tikzpicture}

\begin{axis}[%
width=0.25\linewidth,
height=0.17\linewidth,
scale only axis,
xmin=0,
xmax=600,
ymin=-0.1,
ymax=0.18,
ymajorgrids,
xtick={0,300,600},
xticklabels={0,0.5,1},
ytick={-0.1,0,0.1},
yticklabels={-0.1,0,0.1},
every x tick label/.append style={font=\color{black}, font=\tiny},
every y tick label/.append style={font=\color{black}, font=\tiny},
axis background/.style={fill=white},
legend style={legend cell align=left, align=left, draw=white!15!black}
]
\addplot [color=black, line width=1.0pt]
  table[row sep=crcr]{%
1	0\\
600	0\\
};
\addlegendentry{data1}

\addplot [color=red, line width=2.0pt]
  table[row sep=crcr]{%
1	0\\
300	0\\
301	0.149999999999977\\
600	0.149999999999977\\
};
\addlegendentry{data2}

\addplot [color=blue, line width=1.0pt]
  table[row sep=crcr]{%
1	0.0811845800315041\\
2	0.0803957748431685\\
3	0.0788258286512473\\
4	0.0764899953586564\\
5	0.0734109703765853\\
6	0.0696186701112538\\
7	0.0651499412904286\\
8	0.0600482029536806\\
9	0.0543630245841769\\
10	0.0481496444817822\\
11	0.0414684330576165\\
12	0.0343843062628366\\
13	0.0269660948529236\\
14	0.0192858756142869\\
15	0.0114182710528894\\
18	-0.0125397863225771\\
19	-0.0203854902476905\\
20	-0.028033124779995\\
21	-0.0354083840131807\\
22	-0.0424396084923728\\
23	-0.0490584814713202\\
24	-0.0552006926911872\\
25	-0.0608065632319494\\
26	-0.0658216253636965\\
27	-0.0701971517661377\\
28	-0.0738906289728902\\
29	-0.0768661704399847\\
30	-0.0790948652272618\\
31	-0.080555058902064\\
32	-0.0812325639384426\\
33	-0.0811207975655179\\
34	-0.0802208457270126\\
35	-0.0785414525302031\\
36	-0.076098935286609\\
37	-0.0729170259693319\\
38	-0.0690266406288629\\
39	-0.0644655790066508\\
40	-0.0592781572656804\\
41	-0.0535147774057805\\
42	-0.0472314375477936\\
43	-0.0404891878454237\\
44	-0.0333535373096083\\
45	-0.0258938173103616\\
46	-0.0181825079400824\\
48	-0.00230653592905128\\
50	0.0136588613437425\\
51	0.0214811378547211\\
52	0.0290946994473416\\
53	0.0364255712746626\\
54	0.0434025251585126\\
55	0.0499577716561816\\
56	0.0560276187164845\\
57	0.0615530905231481\\
58	0.0664805005163771\\
59	0.0707619730210354\\
60	0.0743559084165781\\
61	0.0772273873268432\\
62	0.0793485099036388\\
63	0.0806986669070966\\
64	0.0812647399485513\\
65	0.0810412289523583\\
66	0.0800303055948461\\
67	0.0782417922040395\\
68	0.0756930663244475\\
69	0.072408891873124\\
70	0.0684211785282969\\
71	0.0637686716895587\\
72	0.0584965760192517\\
73	0.0526561162254211\\
74	0.04630403935289\\
75	0.0395020634173306\\
76	0.0323162777415291\\
77	0.0248165008183605\\
78	0.01707560194086\\
80	0.00117289865863768\\
82	-0.0147752783436772\\
83	-0.0225726052217397\\
84	-0.030150612271882\\
85	-0.037435670102127\\
86	-0.044356995671933\\
87	-0.0508473400342382\\
88	-0.0568436417395333\\
89	-0.0622876395526646\\
90	-0.0671264385309769\\
91	-0.0713130239611246\\
92	-0.0748067181639271\\
93	-0.0775735757262055\\
94	-0.0795867133211914\\
95	-0.0808265709117677\\
96	-0.081281101800414\\
97	-0.0809458896762862\\
98	-0.0798241915258586\\
99	-0.0779269059867147\\
100	-0.0752724674545107\\
101	-0.0718866669709541\\
102	-0.0678024016328891\\
103	-0.0630593549576588\\
104	-0.0577036113103304\\
105	-0.0517872081391033\\
106	-0.0453676303692419\\
107	-0.0385072518683955\\
108	-0.0312727294100341\\
109	-0.0237343550231799\\
110	-0.0159653730211176\\
112	-3.90331416610934e-05\\
114	0.0158888200672891\\
115	0.0236596799486506\\
116	0.0312006577725015\\
117	0.0384384839297809\\
118	0.0453028342922153\\
119	0.0517270134949968\\
120	0.05764860296199\\
121	0.0630100673770357\\
122	0.0677593137075974\\
123	0.0718501973517505\\
124	0.0752429704872384\\
125	0.0779046682696389\\
126	0.0798094291251346\\
127	0.0809387460261632\\
128	0.0812816463098898\\
129	0.0808347982904252\\
130	0.0796025436301306\\
131	0.0775968551550932\\
132	0.0748372205252963\\
133	0.0713504528880549\\
134	0.0671704303567822\\
135	0.0623377668442799\\
136	0.0568994174502677\\
137	0.0509082222365578\\
138	0.0444223928222982\\
139	0.0375049467893405\\
140	0.0302230953900562\\
141	0.0226475905107009\\
142	0.0148520372318899\\
144	-0.00109483997118787\\
146	-0.0169992698187116\\
147	-0.0247421504902832\\
148	-0.0322446316097285\\
149	-0.0394338176095061\\
150	-0.0462398569586639\\
151	-0.0525966208534783\\
152	-0.0584423457380581\\
153	-0.0637202334112317\\
154	-0.0683790028886051\\
155	-0.0723733886586615\\
156	-0.0756645804915479\\
157	-0.0782206005263788\\
158	-0.0800166139749763\\
159	-0.0810351704208188\\
160	-0.0812663733711361\\
161	-0.0807079764131231\\
162	-0.079365405040221\\
163	-0.0772517039374634\\
164	-0.0743874102363407\\
165	-0.0708003539721176\\
166	-0.0665253876819634\\
167	-0.0616040477708566\\
168	-0.0560841509357033\\
169	-0.0500193295689542\\
170	-0.0434685106561119\\
171	-0.0364953432291486\\
172	-0.029167579940804\\
173	-0.0215564187660675\\
174	-0.0137358112290258\\
178	0.0181064115039362\\
179	0.0258198061973189\\
180	0.0332823306258661\\
181	0.0404214774490583\\
182	0.0471678813263452\\
183	0.0534559928837552\\
184	0.0592247156049552\\
185	0.0644179994568503\\
186	0.0689853854820512\\
187	0.0728824960687007\\
188	0.0760714661313386\\
189	0.0785213110159475\\
190	0.0802082275524754\\
191	0.0811158253314943\\
192	0.0812352859562679\\
193	0.0805654487239735\\
194	0.0791128219036636\\
195	0.0768915195003501\\
196	0.0739231241207108\\
197	0.0702364772724877\\
198	0.0658673991340493\\
199	0.0608583405195304\\
200	0.0552579704178697\\
201	0.0491207031153635\\
202	0.0425061694962778\\
203	0.0354786376573202\\
204	0.028106388466199\\
205	0.0204610521317363\\
206	0.0126169122305555\\
210	-0.0192100296725357\\
211	-0.0268924373573327\\
212	-0.0343135528841003\\
213	-0.0414012712492422\\
214	-0.0480867268011025\\
215	-0.0543049623515799\\
216	-0.0599955603131548\\
217	-0.0651032297281517\\
218	-0.0695783434856594\\
219	-0.0733774205091322\\
220	-0.0764635482264566\\
221	-0.0788067412197506\\
222	-0.0803842325695996\\
223	-0.0811806950628124\\
224	-0.0811883901147894\\
225	-0.0804072429589269\\
226	-0.0788448433730764\\
227	-0.0765163719357815\\
228	-0.0734444525289746\\
229	-0.0696589325198147\\
230	-0.0651965927579568\\
231	-0.060100790205297\\
232	-0.0544210366717834\\
233	-0.048212517748766\\
234	-0.0415355566150311\\
235	-0.0344550279251052\\
236	-0.0270397274746301\\
237	-0.0193617037666627\\
238	-0.0114955579746265\\
242	0.0203099095598418\\
243	0.0279598352358335\\
244	0.0353380977080633\\
245	0.0423730083417695\\
246	0.0489962145752543\\
247	0.0551433640470123\\
248	0.0607547298558302\\
249	0.0657757908788881\\
250	0.0701577615094493\\
251	0.0738580656675367\\
252	0.0768407504775723\\
253	0.079076835593014\\
254	0.0805445947755743\\
255	0.0812297669909867\\
256	0.0811256949727976\\
257	0.0802333899049472\\
258	0.0785615215971802\\
259	0.0761263342476468\\
260	0.0729514886106699\\
261	0.0690678321047926\\
262	0.0645130990928919\\
263	0.0593315442476978\\
264	0.0535735125652081\\
265	0.0472949502025131\\
266	0.0405568608940712\\
267	0.033424713227646\\
268	0.0259678045388227\\
269	0.0182585876045778\\
271	0.00238456970282641\\
273	-0.0135818988593428\\
274	-0.0214058371288957\\
275	-0.0290217921166231\\
276	-0.0363557657209412\\
277	-0.0433364996259797\\
278	-0.0498961676619274\\
279	-0.055971034816821\\
280	-0.0615020764983001\\
281	-0.0664355520284516\\
282	-0.0707235267983606\\
283	-0.0743243380102285\\
284	-0.07720299948096\\
285	-0.0793315415753568\\
286	-0.0806892829638173\\
287	-0.0812630315666638\\
288	-0.0810472127307094\\
289	-0.0800439233938732\\
290	-0.0782629117106808\\
291	-0.0757214823374852\\
292	-0.0724443282967968\\
293	-0.0684632910556502\\
294	-0.063817051146998\\
295	-0.0585507523425122\\
296	-0.0527155630269363\\
297	-0.0463681790357668\\
298	-0.0395702727880689\\
299	-0.0323878940645272\\
300	-0.0248908282554794\\
301	-0.0171519183122655\\
302	-0.0118191447376148\\
303	-0.00814440576175457\\
304	-0.0056121950179886\\
305	-0.00386728434716588\\
306	-0.00266489104069478\\
307	-0.00183633878998535\\
308	-0.0012653951325774\\
309	-0.000871965919486684\\
311	-0.000414043731439051\\
313	-0.000196604256757382\\
317	-4.43288354290416e-05\\
327	-1.07008884242532e-06\\
600	0\\
};
\addlegendentry{data3}

\addplot [color=white!40!black, dotted, line width=2.0pt]
  table[row sep=crcr]{%
1	0.00971619472647944\\
600	0.00971619472647944\\
};
\addlegendentry{data4}
\legend{}
\end{axis}
\end{tikzpicture}%
&
%
%
\begin{tikzpicture}

\begin{axis}[%
width=0.25\linewidth,
height=0.17\linewidth,
scale only axis,
xmin=0,
xmax=600,
ymin=-0.1,
ymax=0.18,
ymajorgrids,
xtick={0,300,600},
xticklabels={0,0.5,1},
ytick={-0.1,0,0.1},
yticklabels={-0.1,0,0.1},
every x tick label/.append style={font=\color{black}, font=\tiny},
every y tick label/.append style={font=\color{black}, font=\tiny},
axis background/.style={fill=white},
legend style={legend cell align=left, align=left, draw=white!15!black}
]
\addplot [color=black, line width=1.0pt]
  table[row sep=crcr]{%
1	0\\
600	0\\
};
\addlegendentry{data1}

\addplot [color=red, line width=2.0pt]
  table[row sep=crcr]{%
1	0\\
300	0\\
301	0.149999999999977\\
600	0.149999999999977\\
};
\addlegendentry{data2}

\addplot [color=blue, line width=1.0pt]
  table[row sep=crcr]{%
1	0.0801487123759443\\
2	0.0727083317183315\\
3	0.0585182772491635\\
4	0.0388958428296746\\
5	0.0156626220597218\\
6	-0.00902459424878543\\
7	-0.0328740376911583\\
8	-0.0536717121539141\\
9	-0.0694869239280251\\
10	-0.0788515122370654\\
11	-0.0808961418208582\\
12	-0.0754310052334404\\
13	-0.0629634430837314\\
14	-0.044650846495415\\
15	-0.0221932139513683\\
16	0.00232466329191539\\
17	0.0266267369460138\\
18	0.0484569922430182\\
19	0.0657888787313823\\
20	0.0770134393965236\\
21	0.0810886736672956\\
22	0.0776362686134462\\
23	0.0669767185639785\\
24	0.0500995729156557\\
25	0.0285715740943715\\
26	0.00439121342026283\\
27	-0.0201967932212028\\
28	-0.042909887276096\\
29	-0.0616395624358574\\
30	-0.0746471018993589\\
31	-0.0807249872656257\\
32	-0.0793089951130241\\
33	-0.0705305750971092\\
34	-0.0552046471904077\\
35	-0.0347539508752561\\
36	-0.0110769691189034\\
37	0.013628311961611\\
38	0.0370684469692151\\
39	0.0570674367899073\\
40	0.0717687313602937\\
41	0.0798075772798938\\
42	0.0804377108606786\\
43	0.073600635421144\\
44	0.0599310516613514\\
45	0.0406979370055751\\
46	0.017686743646891\\
47	-0.00696634889425241\\
48	-0.0309727399971962\\
49	-0.0521038637532456\\
50	-0.0683980716243013\\
51	-0.0783427365764737\\
52	-0.0810146735633452\\
53	-0.0761658408173389\\
54	-0.0642463661110924\\
55	-0.0463627604094654\\
56	-0.0241751980307754\\
58	0.0246645791471565\\
59	0.0467828903732652\\
60	0.064558243336819\\
61	0.076340513058426\\
62	0.0810359256128095\\
63	0.0782085955599996\\
64	0.0681209901464399\\
65	0.0517095638949741\\
66	0.0304978254787329\\
67	0.00645490707131557\\
68	-0.0181872345087868\\
69	-0.0411410152426015\\
70	-0.0602755853495864\\
71	-0.0738146407430804\\
72	-0.0805013212331005\\
73	-0.0797148876125675\\
74	-0.0715283462380967\\
75	-0.0567016716908029\\
76	-0.0366112566690617\\
77	-0.0131221384381206\\
78	0.0115851366680317\\
79	0.0352169381424119\\
80	0.0555794740528199\\
81	0.0707824455554373\\
82	0.0794145274804805\\
83	0.0806743847413145\\
84	0.0744450620264843\\
85	0.0613048410186821\\
86	0.0424735572375994\\
87	0.0196993600351334\\
88	-0.00490357194075841\\
89	-0.0290512945847468\\
90	-0.0505021218712045\\
91	-0.0672647264827901\\
92	-0.0777829990892087\\
93	-0.0810805053881722\\
94	-0.0768511306416713\\
95	-0.0654874969785624\\
96	-0.04804451542077\\
97	-0.0261414561820175\\
98	-0.00181162826311265\\
99	0.0226863770784576\\
100	0.0450783563069308\\
101	0.0632856129066113\\
102	0.0756179273367934\\
103	0.0809304638162303\\
104	0.0787300479382793\\
105	0.0692209491322728\\
106	0.0532859178845229\\
107	0.0324042380756282\\
108	0.00851440181543239\\
109	-0.0161658450290361\\
110	-0.0393453810435176\\
111	-0.0588723990898643\\
112	-0.0729341632792284\\
113	-0.0802252892179922\\
114	-0.0800689257032445\\
115	-0.072479588302599\\
116	-0.0581618118429788\\
117	-0.038444746898108\\
118	-0.0151587718269184\\
119	0.00953442526122217\\
120	0.0333425207527398\\
121	0.0540553569553595\\
122	0.0697501158812202\\
123	0.0789698186561054\\
124	0.0808585800615447\\
125	0.0752410622362731\\
126	0.0626387516734894\\
127	0.0442215484873714\\
128	0.0216991620206954\\
129	-0.00283760522142984\\
130	-0.0271109513507781\\
131	-0.0488675284384499\\
132	-0.0660876257425116\\
133	-0.0771726638834025\\
134	-0.0810935944718949\\
135	-0.0774864289265906\\
136	-0.0666860283322421\\
137	-0.0496950175505617\\
138	-0.0280907093571159\\
139	-0.00387867903862116\\
140	0.0206934175572542\\
141	0.0433444988407246\\
142	0.0619718152848918\\
143	0.074846152272471\\
144	0.0807723568804022\\
145	0.0792002865442782\\
146	0.0702758799994854\\
147	0.0548276094682478\\
148	0.0342895717667488\\
149	0.0105683579546394\\
150	-0.0141339396924423\\
151	-0.0375241527360686\\
152	-0.0574309164273927\\
153	-0.0720062422569754\\
154	-0.0798970707785429\\
155	-0.0803708790837163\\
156	-0.0733836825088474\\
157	-0.0595841178281944\\
158	-0.0402532288798056\\
159	-0.0171855444585844\\
160	0.00747751172582412\\
161	0.0314464141058579\\
162	0.0524960769332665\\
163	0.0686724138663521\\
164	0.0784737400889526\\
165	0.0809901770028318\\
166	0.0759881182534627\\
167	0.0639319159189427\\
168	0.0459407736896082\\
169	0.0236848487354564\\
171	-0.0251529724857846\\
172	-0.0472011467553557\\
173	-0.0648675351058046\\
174	-0.0765121279813457\\
175	-0.0810539323000512\\
176	-0.0780713224115743\\
177	-0.0678411805295127\\
178	-0.0513131931495536\\
179	-0.030021689569935\\
180	-0.00594320673940274\\
181	0.0186869970003727\\
182	0.0415824458464158\\
183	0.0606177050948418\\
184	0.0740256899040332\\
185	0.0805617076534872\\
186	0.0796190054885528\\
187	0.0712850965176131\\
188	0.056333635778401\\
189	0.0361526001453285\\
190	0.0126154393935849\\
191	-0.0120928402504887\\
192	-0.035678515026234\\
193	-0.055952075044047\\
194	-0.0710314812874913\\
195	-0.0795168794203391\\
196	-0.0806205513340501\\
197	-0.0742400407452806\\
198	-0.0609676644385218\\
199	-0.04203552619947\\
200	-0.0192011379205042\\
201	0.00541573408088425\\
202	0.0295298516161893\\
203	0.0509026482958461\\
204	0.0675500405544653\\
205	0.0779266144774056\\
206	0.0810690899614883\\
207	0.0766857441193451\\
208	0.0651834925538424\\
209	0.0476301144908575\\
210	0.025655128493554\\
211	0.00129852068153014\\
212	-0.0231786316517173\\
213	-0.0455040608006811\\
214	-0.0636052482395826\\
215	-0.075801821060395\\
216	-0.0809615446729595\\
217	-0.0786054306241795\\
218	-0.0689522021456241\\
219	-0.052897989597227\\
220	-0.0319331407210939\\
221	-0.00800386839364364\\
222	0.0166684205814818\\
223	0.0397933435364166\\
224	0.059224163183103\\
225	0.0731570739408198\\
226	0.0802986531626857\\
227	0.079985932395175\\
228	0.0722479421925755\\
229	0.05780301714708\\
230	0.0379921113144519\\
231	0.0146543145090163\\
232	-0.0100438744349276\\
233	-0.033809668498634\\
234	-0.0544368369240829\\
235	-0.070010514451269\\
236	-0.0790849624572729\\
237	-0.0808177800425938\\
238	-0.0750481059521917\\
239	-0.0623115516791586\\
240	-0.0437904794755468\\
241	-0.0212042410724962\\
242	0.00335043350946762\\
243	0.0275940800044054\\
244	0.0492761075657882\\
245	0.0663837260475475\\
246	0.0773287977258406\\
247	0.0810952676048373\\
248	0.0773334860294881\\
249	0.0663926674293407\\
250	0.0492884719776612\\
251	0.0276087196308481\\
252	0.00336598932187826\\
253	-0.0211892131542299\\
254	-0.0437773745260301\\
255	-0.0623015862596503\\
256	-0.0750422051743271\\
257	-0.0808164916885517\\
258	-0.0790884061278803\\
259	-0.0700183704628898\\
260	-0.0544483759859986\\
261	-0.0338238194144651\\
262	-0.0100593235440556\\
263	0.0146390013811697\\
264	0.0379783557192468\\
265	0.0577920960464553\\
266	0.0722408694161913\\
267	0.0799833645245371\\
268	0.0803008285787428\\
269	0.0731637906949345\\
270	0.0592347977442387\\
271	0.039806908675132\\
272	0.0166836570159603\\
273	-0.0079883750947829\\
274	-0.0319188288348187\\
275	-0.0528861877274949\\
276	-0.068944005885669\\
277	-0.0786016008506749\\
278	-0.0809624369123867\\
279	-0.075807352484162\\
280	-0.0636149053534609\\
281	-0.0455169471135832\\
282	-0.0231935508991228\\
283	0.00128295348622487\\
284	0.0256403584870668\\
285	0.0476175128054592\\
286	0.0651742290314132\\
287	0.07668067871316\\
288	0.0810686929041822\\
289	0.0779309226286387\\
290	0.0675586539792903\\
291	0.0509147673909638\\
292	0.0295443513394957\\
293	0.00543126839136221\\
294	-0.0191860111069673\\
295	-0.0420222111381463\\
296	-0.0609573971958071\\
297	-0.0742337744518409\\
298	-0.0806188677034925\\
299	-0.0795199347479638\\
300	-0.0710389919403269\\
301	-0.0559633437919729\\
302	-0.0440869973352847\\
303	-0.0347310078766441\\
304	-0.0273605140072277\\
305	-0.0215541607486784\\
306	-0.0169800116131\\
307	-0.0133765725207695\\
308	-0.0105378427577989\\
309	-0.00830153836602676\\
310	-0.00653981472555643\\
311	-0.00515195795753698\\
312	-0.00405862733271078\\
313	-0.00319731953595692\\
314	-0.00251879548841316\\
315	-0.001984265457736\\
317	-0.00123144069982573\\
319	-0.000764235546853342\\
322	-0.000373635239952819\\
326	-0.000143904804076556\\
333	-2.70971335112336e-05\\
356	-1.122745061366e-07\\
600	0\\
};
\addlegendentry{data3}

\addplot [color=white!40!black, dotted, line width=2.0pt]
  table[row sep=crcr]{%
1	0.0928321920213193\\
600	0.0928321920213193\\
};
\addlegendentry{data4}
\legend{}
\end{axis}
\end{tikzpicture}%
&
  \input{./figures/perc3.tex}
  \end{tabular}
  \begin{overpic}
  [trim=0cm -1.8cm 0cm 0cm,clip,width=1.0\linewidth]{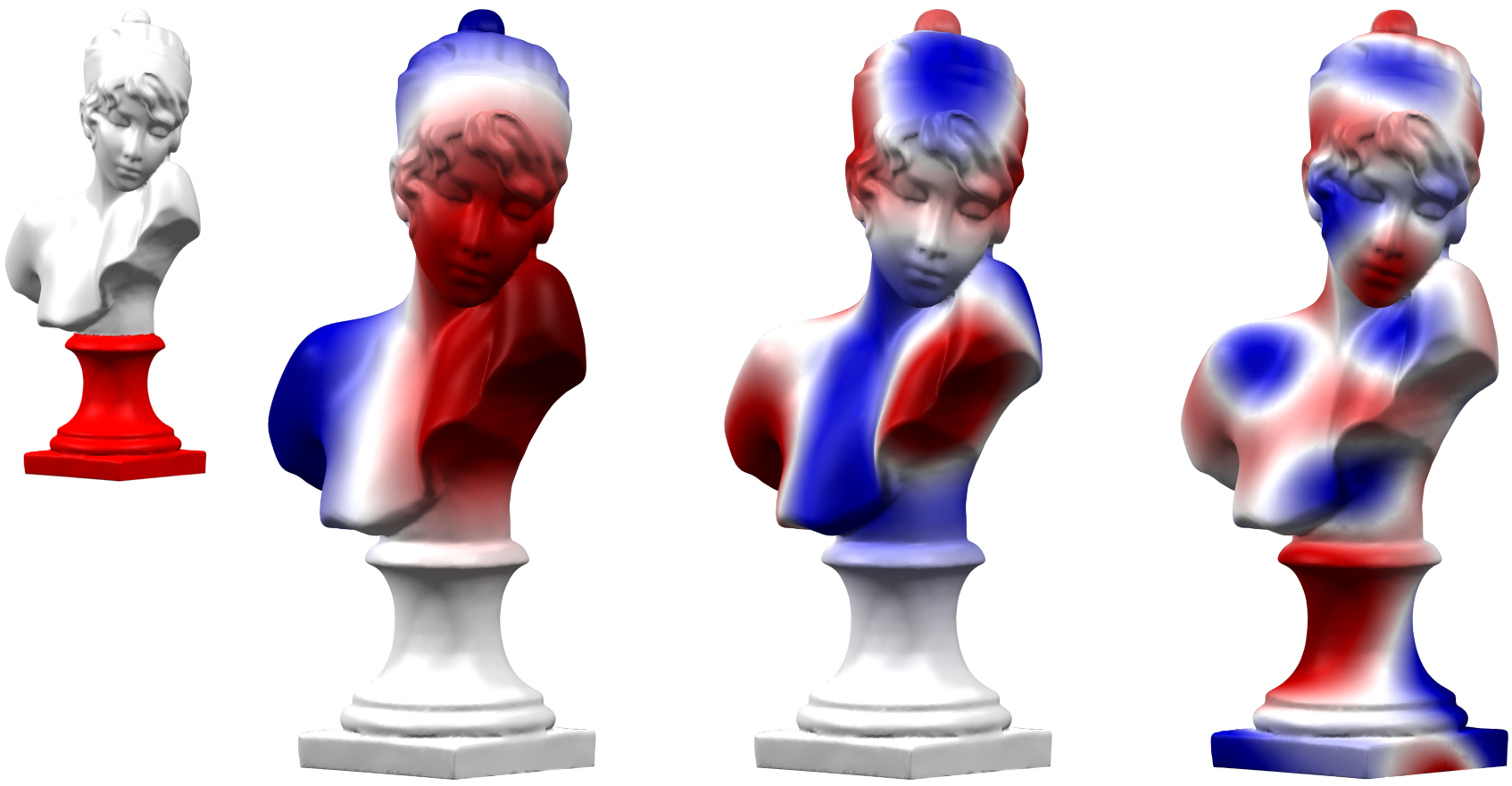}
  \end{overpic}
\caption{\label{fig:stepwall}{\em Top}: A step potential $v_\tau$ (solid red) confines eigenfunction $\psi_i$ (solid blue) within the region where $v_\tau(x)=0$, as long as for the associated eigenvalue $\lambda_i$ (dashed gray) it holds $\lambda_i<\tau$. As soon as $\lambda_i \ge \tau$, a ``diffusion'' effect across the potential barrier takes place (rightmost plot). We illustrate this in $\mathbb{R}$ with three eigenfunctions at increasing eigenvalue (left to right). {\em Bottom}: We show the same phenomenon on a surface. Here, the potential with height $\tau$ is depicted on the left and is supported on the pedestal; observe the diffusion in the rightmost figure, due to the eigenvalue of the plotted eigenfunction being higher than $\tau$.}
\end{figure}

A classical result in mathematical physics \cite[Ch.~2]{griffiths} shows that the eigenfunctions $\psi_i$ of $H_\X$ exhibit {\em localization} depending not only on the geometry of the domain $\X$ \cite{heilman2010localized}, but also on the shape of the potential $v$.
In this paper, we consider step potentials with finite height $\tau>0$:
\begin{align}\label{eq:step}
v_\tau(x) = \left\{ \begin{array}{ll}
         0 & x\in \R\\
         \tau  &  x\notin\R\end{array} \right.
\end{align}
for a given region $\R\subseteq\X$.
The following lemma on the ``confinement'' property of step potentials is instrumental to our approach: 

  \begin{lemma}[\!{\!\cite[Ch.~2.6]{griffiths}}]
    Let $\R\subseteq\X$ be a region of $\X$, and let $v_\tau:\X\to\{0,\tau\}$ be a finite step potential defined as in Eq.~\ref{eq:step}. Then, for the Hamiltonian $\Delta_\X+v_\tau$ all eigenfunctions $\psi_i$ with eigenvalue $\lambda_i<\tau$ vanish pointwise for $x\notin\R$.
  \label{thm:step}
  \end{lemma}

According to Lemma~\ref{thm:step}, for large enough $\tau$, the Hamiltonian eigenfunctions will be entirely supported within $\R$. This result holds both on the real line (where the eigenfunctions are Fourier harmonics) and on manifolds\footnote{Berger \cite[p.403]{berger2012panoramic} described this analogy quite vividly, by depicting manifolds as ``quantum mechanical worlds''.}. We illustrate real examples of these facts in Figure \ref{fig:stepwall}. 

Key to our approach is the realization that the confined behavior or Hamiltonian eigenfunctions can be exploited by optimizing over the space of potentials $v_\tau$. We therefore establish the following lemma:

  \begin{lemma}[Eigenvalue equivalence]
    Let $\R\subseteq\X$ be a region of $\X$, and let $v_\tau:\X\to\{0,\tau\}$ be a finite step potential defined as in Eq.~\eqref{eq:step}. Now let $\{(\lambda_i,\psi_i)\}_{i=1}^{k}$ be $k$ eigenpairs of $\Delta_\X+v_\tau$, increasingly ordered and such that $\lambda_k<\tau$. If $(\mu_i,\phi_i)$ are the eigenpairs of $\Delta_{\R}$ computed with Dirichlet boundary conditions, then $\mu_i=\lambda_i$ for all $i=1,\dots,k$.
  \label{thm:diri}
  \end{lemma}

\noindent\textbf{Sketch of proof.}
For any eigenpair $(\lambda_i,\psi_i)$ of $\Delta_\X+v_\tau$ with $\lambda_i<\tau$, the following equality holds pointwise:
\begin{align}
    \lambda_i \psi_i(x) &= \Delta_\X\psi_i(x) + v_\tau(x) \psi_i(x) \label{eq:l1} \\
    &= \Delta_{\R}\psi_i(x) \label{eq:l2} \\
    &= \mu_i\phi_i(x) \label{eq:l3} \,.
\end{align}
The equivalence~\eqref{eq:l2} follows directly from Lemma~\ref{thm:step}, since $\psi_i(x)$ vanishes for $x\notin\R$. Eq.~\eqref{eq:l3} stems from the homogeneous Dirichlet condition $\phi_i(x)=0$ for $x\in\partial\R$. \hfill $\square$

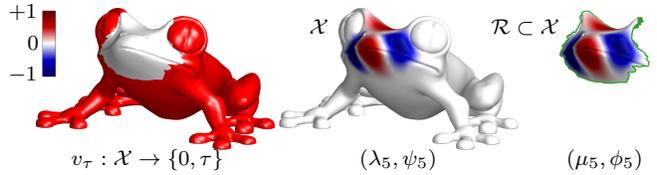
\begin{figure}[t]
  \centering
\begin{overpic}
[trim=0cm -1.8cm 0cm 0cm,clip,width=1.0\linewidth]{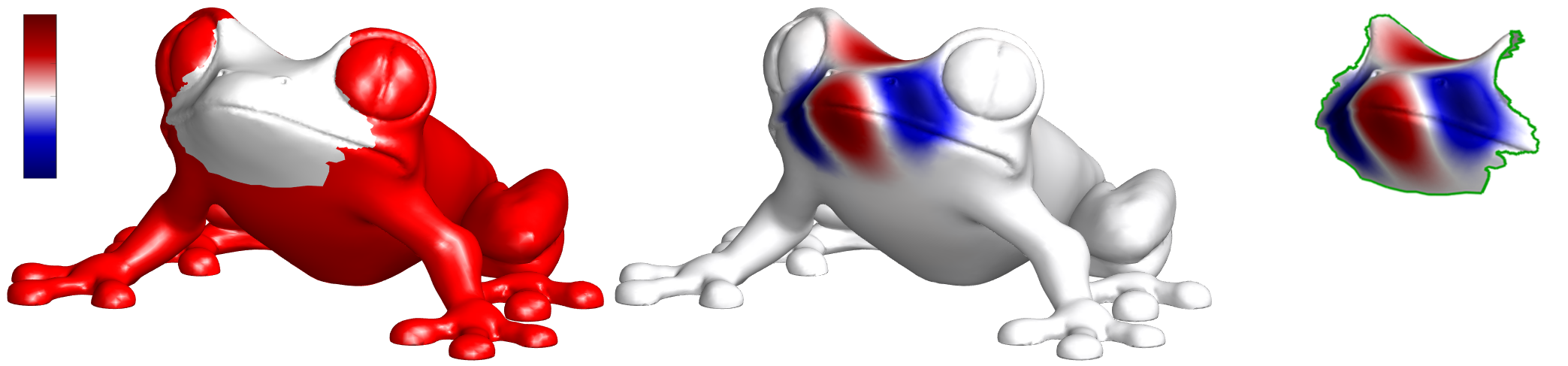}
\put(-4,24.5){\footnotesize $+1$}
\put(-1,19.5){\footnotesize $0$}
\put(-4,14.5){\footnotesize $-1$}
\put(6,1){\footnotesize $v_\tau:\X\to\{0,\tau\}$}
\put(44,22){\footnotesize $\X$}
\put(73,22){\footnotesize $\R\subset\X$}
\put(52,1){\footnotesize $(\lambda_5,\psi_5)$}
\put(85,1){\footnotesize $(\mu_5,\phi_5)$}
\end{overpic}
  \caption{\label{fig:dirichlet}{\em Left}: Given a region $\R$ (the face of the frog), we show the step potential $v_\tau$ such that $v_\tau(x)=0$ for $x\in\R$, and $v_\tau(x)=\tau$ otherwise. {\em Middle}: The eigenfunctions $\psi_i$ of the Hamiltonian $\Delta_\X+v_\tau$ are localized on $\R$, meaning that $\psi_i(x)\approx 0$ for $x\notin\R$. {\em Right}: If the region $\R$ is taken as a separate mesh with boundary $\partial\R$ (depicted in green), the Dirichlet eigenfunction $\phi_i$ of $\Delta_\R$ has the same values as $\psi_i$ at corresponding points. As a consequence, $\lambda_i=\mu_i$.}
\end{figure}

\vspace{1ex}
An illustration of this result is given in Figure~\ref{fig:dirichlet}.
Lemma~\ref{thm:diri} establishes that one can measure the Dirichlet eigenvalues of a partial shape $\R$ equivalently by measuring Hamiltonian eigenvalues over the complete shape $\X$ -- provided that an indicator for the corresponding region on $\X$ is given. This gives us a straightforward way to implement localization via optimization over the cone of non-negative real-valued functions on $\X$. Problem~\eqref{eq:prob} in the next Section formalizes precisely this idea.

%

\section{Our method}\label{sec:method}
\vspace{-0.5ex}
We first present and discuss our general formulation, and then demonstrate its application to two separate tasks in Sections~\ref{sec:local} and \ref{sec:isosp}.
The main idea behind our approach is to model regions $\R\subset\X$ via potential functions, rather than by explicitly manipulating points $x\in\R$.
Further, similarly to \cite{isosp}, we rely upon the empirical observation that, in many practical cases, the Laplacian eigenvalues encode the geometry of the domain.

\vspace{1ex}\noindent\textbf{Optimization problem.}
Our input is a given shape $\X$ and, separately, an increasing sequence of $k$ Dirichlet eigenvalues stored in a vector $\bm{\mu}\in\mathbb{R}^{k}$. The precise source of these eigenvalues is {\em not known}, but they are assumed to come from a manifold Laplacian. It is our task to identify the shape of this domain.

To do so, we consider the following {\em eigenvalue alignment} problem:
\begin{align}\label{eq:prob}
\min_{\v\ge 0} \|  \bm{\lambda}(\bm{\Delta}_\X + \mathrm{diag}(\v)) - \bm{\mu} \|_w^2\,,
\end{align}
where we seek for an optimal alignment (according to a weighted norm defined below) between the input eigenvalues $\bm{\mu}$ and the Hamiltonian eigenvalues $\bm{\lambda}$; the minimization is carried out over the space of non-negative potentials on $\X$.
Problem \eqref{eq:prob} therefore models the input sequence $\bm{\mu}$ as the first $k$ eigenvalues of a Hamiltonian (whose potential $v$ we seek) constructed on $\X$. This allows us to model very general tasks in practice, as we show in our experiments.

The $w$-norm used in Eq.~\eqref{eq:prob} is a weighted $L_2$ norm:
\begin{align}
\| \bm{\lambda} - \bm{\mu} \|_w^2 = \sum_{i=1}^k \frac{1}{\mu_i^2}(\lambda_i - \mu_i)^2\,,
\end{align}
where the weight $\frac{1}{\mu_i^2}$ balances the contribution of each term, thus avoiding unduly penalization of lower frequencies.

\vspace{1ex}\noindent\textbf{Remark.}
We emphasize that our objective is phrased entirely in terms of eigenvalues, while the eigenfunctions never appear explicitly. In the applications, we will show a remarkable property where alignment of eigenvalues also promotes alignment of the associated eigenfunctions.


\vspace{1ex}\noindent\textbf{Algorithm.}
Although the optimization problem in Eq. \eqref{eq:prob} is non-convex, it is still differentiable and we observed very good local optima by using standard numerical optimization methods. For example, projected gradient descent with step $\alpha>0$ can be easily implemented by the recursive relations:
\begin{align*}
\v^{(t)} \hspace{-0.1cm}&= \v^{(t-1)} \hspace{-0.065cm}-\hspace{-0.005cm} \Pi\left(\alpha \nabla \| \bm{\lambda}(\bm{\Delta}+\mathrm{diag}(\v^{(t-1)}))-\bm{\mu}\|^2_w\right)\nonumber\\
&=\v^{(t-1)} \hspace{-0.065cm}- \hspace{-0.005cm} \Pi\left( 2\alpha ( \bm{\Phi}^{(t-1)} \hspace{-0.04cm} \circ \bm{\Phi}^{(t-1)} ) ((\bm{\lambda}^{(t-1)}\hspace{-0.075cm}-\bm{\mu}){\scriptscriptstyle\oslash}\bm{\mu}^2) \right)
\end{align*}
Here, $\Pi(\mathbf{x})\equiv\max\{\mathbf{x},0\}$, with element-wise $\max$, is a projector onto $\mathbb{R}^n_+$; element-wise product and division are denoted by $\circ$ and $\scriptscriptstyle\oslash$ respectively. The matrix $\bm{\Phi}^{(t-1)}$ contains the eigenvectors of $\bm{\Delta}+\mathrm{diag}(\v^{(t-1)})$ as its columns; note that we do {\em not} optimize over these eigenvectors, but they appear naturally in the eigenvalue derivatives \cite{deigs}.

To simplify even further, we consider the {\em unconstrained} problem
\begin{align}\label{eq:prob2}
\min_{\v\in\mathbb{R}^n} \| \bm{\lambda}(\bm{\Delta}_\X + \mathrm{diag}(\sigma(\v))) - \bm{\mu} \|_w^2\,,
\end{align}
where $\sigma(\v)= \frac{\tau}{2} (\tanh(\v) + 1)$ is a saturation function, acting element-wise, which maps the values of $\v$ to within the range $(0,\tau)$. This has the effect of promoting step potentials with height $\tau$, ensuring that all eigenfunctions with associated eigenvalue $\lambda_i<\tau$ will be confined within the region where $\sigma(v(x)) = 0$ (by Lemma~\ref{thm:step}). In our tests we simply set $\tau = 10 \, \mu_k $. We minimize problem~\eqref{eq:prob2} by a standard trust-region method implemented in Matlab. 



%
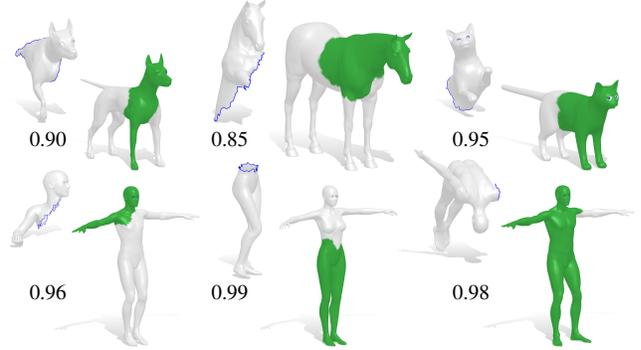
\begin{figure}[t]
  \centering
\begin{overpic}
[trim=0cm 0cm 0cm 0cm,clip,width=1.0\linewidth]{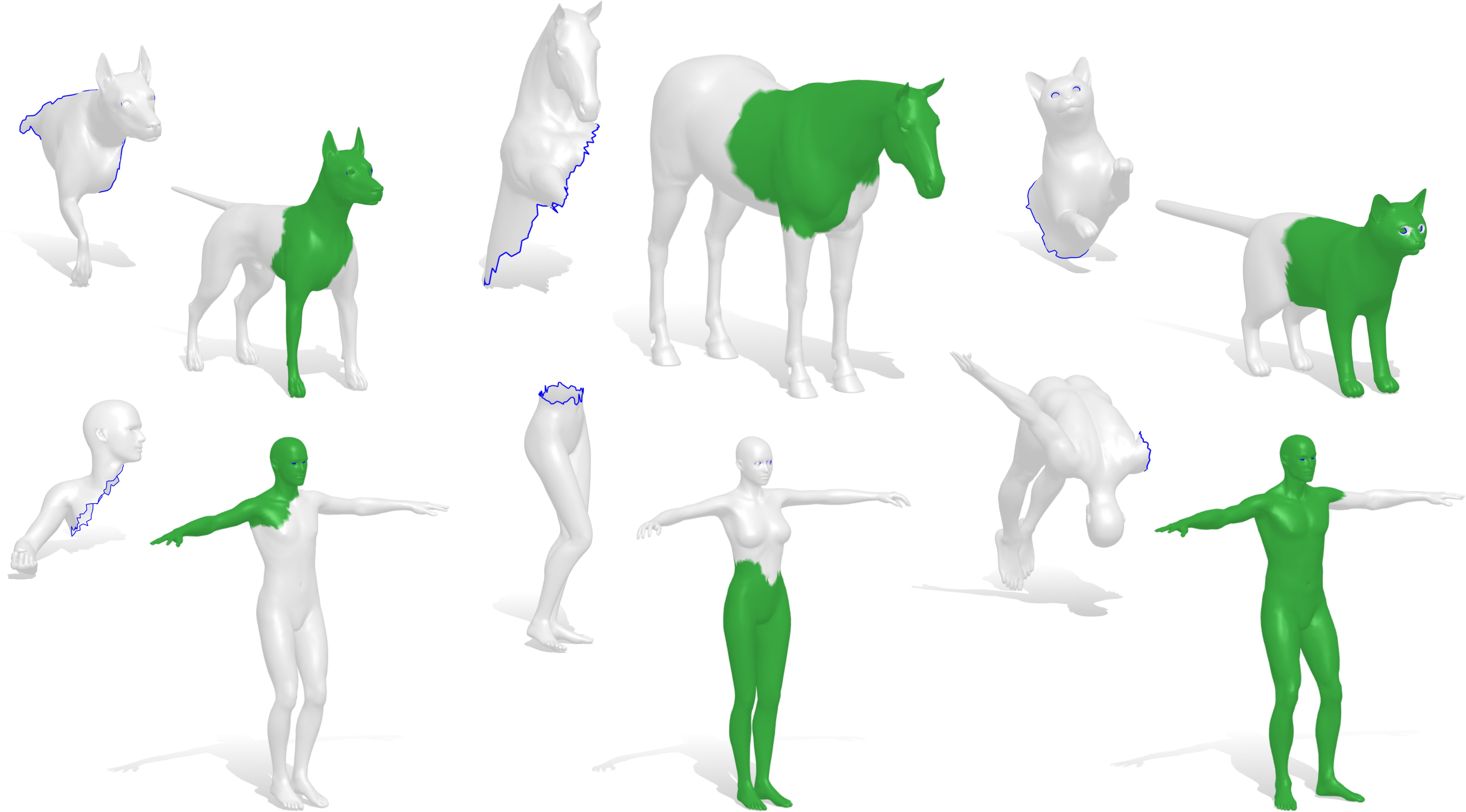}
\put(3.5,32){\footnotesize 0.90}
\put(32.5,32){\footnotesize 0.85}
\put(71,32){\footnotesize 0.95}
\put(3.5,7.5){\footnotesize 0.96}
\put(32.5,7.5){\footnotesize 0.99}
\put(71,7.5){\footnotesize 0.98}
\end{overpic}
  \caption{\label{fig:examples}Examples of correctly detected regions using our method on 6 different classes from the SHREC'16 dataset \cite{shrec16}. For each example we show the partial 3D query and its optimal localization on the full model (in green). Below each partial shape we also report the intersection-over-union area ratio between the detected region and the ground truth (ideally 1.0).}
\end{figure}

\section{Results}
\vspace{-0.5ex}
We showcase our eigenvalue alignment method on two distinct tasks. As we present these results, we also highlight some key properties of our technique that can be exploited in several applications.

\subsection{Partial shape localization}\label{sec:local}
As a first application, we consider the task of finding a region of a non-rigid shape from a partial (and possibly deformed) 3D query. Differently from prior work, we aim to do this without, at the same time, having to compute a map between full and partial shape.

Let $\X$ and $\Y$ be the full and partial shapes, respectively, with the Laplacians $\Delta_\X$ and $\Delta_\Y$. Further, let $\{\mu_i\}_{i=1}^{k}$ be the first $k$ eigenvalues of $\Delta_\Y$. Our task is to find a region $\R\subseteq\X$ that corresponds to the partial query $\Y$ up to isometry.

To solve this problem, we minimize Eq.~\eqref{eq:prob2} directly by using $\Delta_\X$ and $\{\mu_i\}$ as input. A local solution will be a potential function $v_\tau:\X\to(0,\tau)$ indicating the sought region $\R\subseteq\X$; see Figure~\ref{fig:examples} for qualitative examples.

\begin{property}
Our method is \textbf{correspondence-free}, since we never need to invoke the notion of a map $\pi:\X\to\Y$ in order to find the region $\R\subseteq\X$. In particular, we only operate with eigenvalue sequences $\{\lambda_i\},\{\mu_i\}$, which are invariant to point ordering (see Eq.~\eqref{eq:de}).
\end{property}

\begin{figure*}[t]
  \centering
\begin{overpic}
[trim=0cm 0cm 0cm 0cm,clip,width=1.0\linewidth]{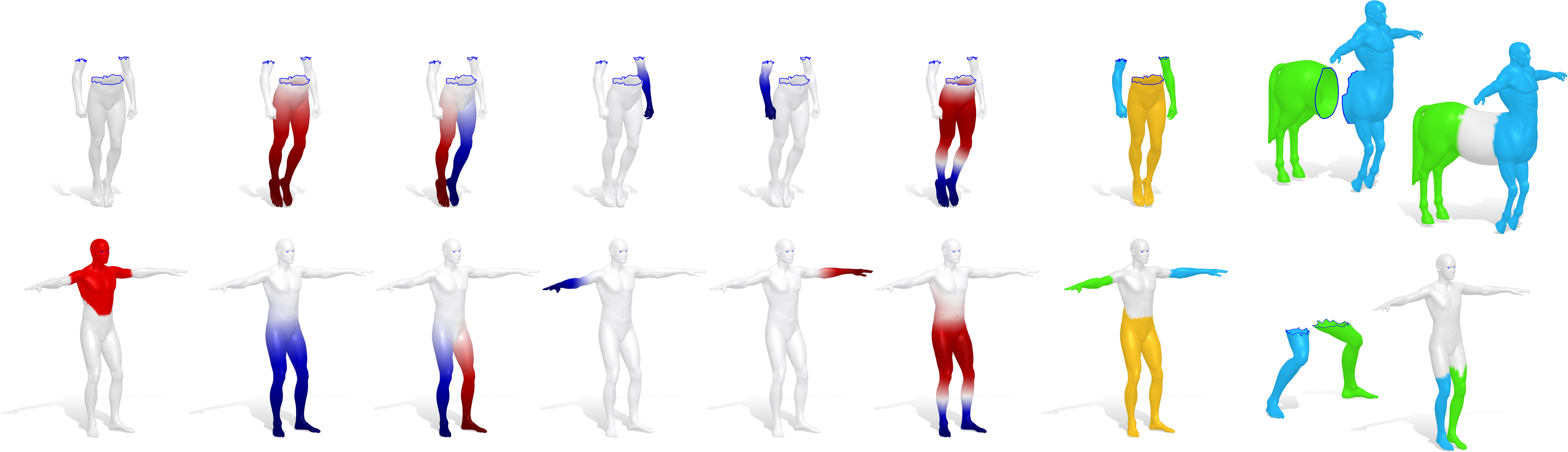}
\put(2.5,20.5){\footnotesize $\Y$}
\put(2.5,7){\footnotesize $\X$}
\put(5.5,10){\footnotesize $\color{white} v_\tau$}
\put(20,25.5){\footnotesize $\phi_1$}
\put(20,13){\footnotesize $\psi_1$}
\put(30.5,25.5){\footnotesize $\phi_2$}
\put(30.5,13){\footnotesize $\psi_2$}
\put(41,25.5){\footnotesize $\phi_3$}
\put(41,13){\footnotesize $\psi_3$}
\put(51.5,25.5){\footnotesize $\phi_4$}
\put(51.5,13){\footnotesize $\psi_4$}
\put(62,25.5){\footnotesize $\phi_5$}
\put(62,13){\footnotesize $\psi_5$}
\put(79,4){\line(0,1){18}}
\end{overpic}
  \caption{\label{fig:puzzles}Simultaneous localization of multiple disconnected regions. Given a query shape $\Y$ (here consisting of 3 pieces), our method solves for a potential $v_{\tau}: \X \to (0, \tau)$ which isolates the regions on $\X$ corresponding to $\Y$ up to isometry (leftmost column). By Property~\ref{thm:align}, the found $v_\tau$ induces alignment of the eigenfunctions (intermediate columns); we exploit this to easily associate a different label to each region. On the right we plot the identified regions with different colors. The rightmost column shows additional multi-piece examples.}
\end{figure*}

Given the property above, it is remarkable to observe that, at least empirically, aligning the (Hamiltonian) eigenvalues of $\X$ to the eigenvalues of $\Y$ provides a very strong prior that also induces an alignment of the associated {\em eigenfunctions}. 
%
 This surprising property, illustrated in Figures~\ref{fig:puzzles} and  \ref{fig:star}, is summarized below.

\begin{property}
Empirically, \textbf{eigenfunction alignment} is induced by our eigenvalue alignment. Formally, this means that $\lambda_i=\mu_i$ implies $\psi_i(x)=\phi_i(\pi(x))$ for all $x\in\R$ and an isometric map $\pi:\R\to\Y$.
\label{thm:align}
\end{property}

We stress that the map $\pi$ is {\em not} solved for, and is not available to us after optimization. 
A similar property was also observed in \cite{isosp}, and might be a feature common to eigenvalue alignment approaches. A deeper comparison to \cite{isosp} will be provided in Section~\ref{sec:isosp}.
Finally, invariance to deformations is put in evidence below:

\begin{property}
Since the eigenvalues of $\Delta_\X$ and $\Delta_\Y$ are \textbf{isometry-invariant}, so are all solutions to problem~\eqref{eq:prob2}.
\end{property}

The latter property suggests that other isometry-invariant methods from the shape analysis area might also be used for the same task. We now discuss the most relevant methods.

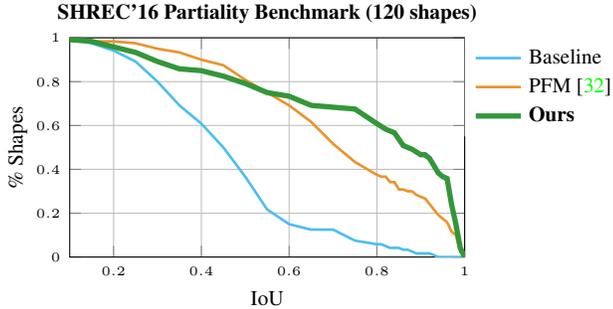
\begin{figure}[t]
  \centering
%
%
\definecolor{mycolor1}{rgb}{0.00000,0.44700,0.74100}%
\definecolor{mycolor2}{rgb}{0.85000,0.32500,0.09800}%
\definecolor{mycolor3}{rgb}{0.92900,0.69400,0.12500}%
\definecolor{mycolor4}{rgb}{0.49400,0.18400,0.55600}%
\definecolor{mycolor5}{rgb}{0.2,0.5882,0.2235}%
\definecolor{mycolor6}{rgb}{0.30100,0.74500,0.93300}%
\definecolor{mycolor7}{rgb}{0.63500,0.07800,0.18400}%
\definecolor{mycolor8}{rgb}{0.9216,0.5765,0.1725}%
\begin{tikzpicture}

\begin{axis}[%
width=0.63\linewidth,
height=0.35\linewidth,
scale only axis,
xmin=0.1,
xmax=1,
ymin=0,
ymax=1,
ymajorgrids,
xmajorgrids,
every x tick label/.append style={font=\color{black}, font=\tiny},
every y tick label/.append style={font=\color{black}, font=\tiny},
axis background/.style={fill=white},
xlabel={\footnotesize IoU},
xlabel style={font=\small, at={(axis description cs:0.5,0.07)},anchor=north},
ylabel={\footnotesize \% Shapes},
ylabel style={font=\small, at={(axis description cs:0.055,0.5)},anchor=north},
title={\footnotesize \textbf{SHREC'16 Partiality Benchmark (120 shapes)}},
title style={at={(0.5,0.95)}},
legend style={legend cell align=left, align=left, draw=white!15!white, at={(1,1)}, anchor=north west, font=\footnotesize}
]



\addplot [color=mycolor6, line width=1.0pt]
  table[row sep=crcr]{%
0.1	1\\
0.15	0.975\\
0.2	0.941666666666667\\
0.25	0.891666666666667\\
0.3	0.8\\
0.35	0.691666666666667\\
0.4	0.608333333333333\\
0.45	0.5\\
0.5	0.366666666666667\\
0.55	0.216666666666667\\
0.6	0.15\\
0.65	0.125\\
0.7	0.125\\
0.75	0.075\\
0.8	0.0583333333333333\\
0.81	0.0583333333333333\\
0.83	0.0416666666666667\\
0.85	0.0416666666666667\\
0.86	0.0333333333333334\\
0.87	0.0333333333333334\\
0.89	0.0166666666666666\\
0.92	0.0166666666666666\\
0.94	0\\
1	0\\
};
\addlegendentry{Baseline}

\addplot [color=mycolor8, line width=1.0pt]
  table[row sep=crcr]{%
0.1	0.991666666666667\\
0.15	0.983333333333333\\
0.2	0.983333333333333\\
0.25	0.975\\
0.3	0.95\\
0.35	0.933333333333333\\
0.4	0.9\\
0.45	0.875\\
0.5	0.808333333333333\\
0.6	0.691666666666667\\
0.65	0.616666666666667\\
0.7	0.516666666666667\\
0.75	0.433333333333333\\
0.8	0.375\\
0.81	0.366666666666667\\
0.82	0.366666666666667\\
0.83	0.341666666666667\\
0.84	0.341666666666667\\
0.85	0.308333333333333\\
0.86	0.308333333333333\\
0.87	0.3\\
0.88	0.3\\
0.89	0.283333333333333\\
0.91	0.266666666666667\\
0.94	0.191666666666667\\
0.96	0.158333333333333\\
0.97	0.116666666666667\\
0.98	0.1\\
0.99	0.0249999999999999\\
1	0\\
};
\addlegendentry{PFM \cite{rodola2017partial}}


\addplot [color=mycolor5, line width=2.0pt]
  table[row sep=crcr]{%
0.1	0.991666666666667\\
0.15	0.983333333333333\\
0.25	0.933333333333333\\
0.3	0.891666666666667\\
0.35	0.858333333333333\\
0.4	0.85\\
0.45	0.825\\
0.5	0.791666666666667\\
0.55	0.75\\
0.6	0.733333333333333\\
0.65	0.691666666666667\\
0.75	0.675\\
0.8	0.608333333333333\\
0.82	0.583333333333333\\
0.84	0.566666666666667\\
0.86	0.508333333333333\\
0.88	0.491666666666667\\
0.9	0.466666666666667\\
0.91	0.466666666666667\\
0.92	0.45\\
0.94	0.383333333333333\\
0.95	0.366666666666667\\
0.96	0.358333333333333\\
0.97	0.241666666666667\\
0.98	0.15\\
0.99	0.0416666666666667\\
1	0\\
};
\addlegendentry{\textbf{Ours}}


\end{axis}
\end{tikzpicture}%
\caption{\label{fig:iou}Partial similarity comparisons. We compare with partial functional maps (PFM), a state-of-the-art partial shape matching method. We measure the intersection-over-union (IoU) between the regions identified by each method and the ground truth regions, and plot the cumulative scores over a set of 120 shape pairs.}
\end{figure}



\vspace{1ex}\noindent\textbf{Comparisons.}
%
Partial shape matching methods such as \cite{bronstein2009partial,rodola2017partial,litany2016non} iteratively compute a map and a region. 
 Such methods make use of regularization functionals for the region in order to reach a good solution. While our potential $v_\tau$ plays a similar role as the membership functions of these methods, we neither use any region regularizer, nor do we need to solve for a map. The fully spectral approach of Litany~\etal~\cite{litany2017fully} seeks for an optimal alignment between {\em eigenfunctions} rather than eigenvalues, thereby solving a pointwise matching problem in the process.

We investigated extensively the effectiveness of our method in comparison with the state-of-the-art shape matching approach partial functional maps (PFM) \cite{rodola2017partial} on the SHREC'16 Partiality benchmark \cite{shrec16}, where PFM showed top performance.
The dataset consists of 120 partial deformable shapes with large missing parts (`cuts' challenge), belonging to 8 sub-classes of humans and animals. The challenge is to locate each of the 120 partial shapes within a full template of the respective class (\eg, a dog in neutral pose). To make the setting even more challenging, we remeshed all the full models $\X$ to half of their density ($\sim$10$K$ to $\sim$5$K$ vertices) via edge collapse \cite{garland1997surface}, while keeping the partial shapes $\Y$ at their original resolution. This way, we eliminate the possibility for any method to work well by virtue of similar  tessellation.

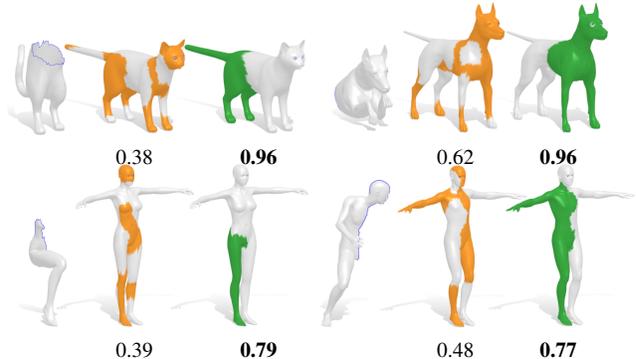
\begin{figure}[t]
  \centering
\begin{overpic}
[trim=0cm -0.5cm 0cm 0cm,clip,width=1.0\linewidth]{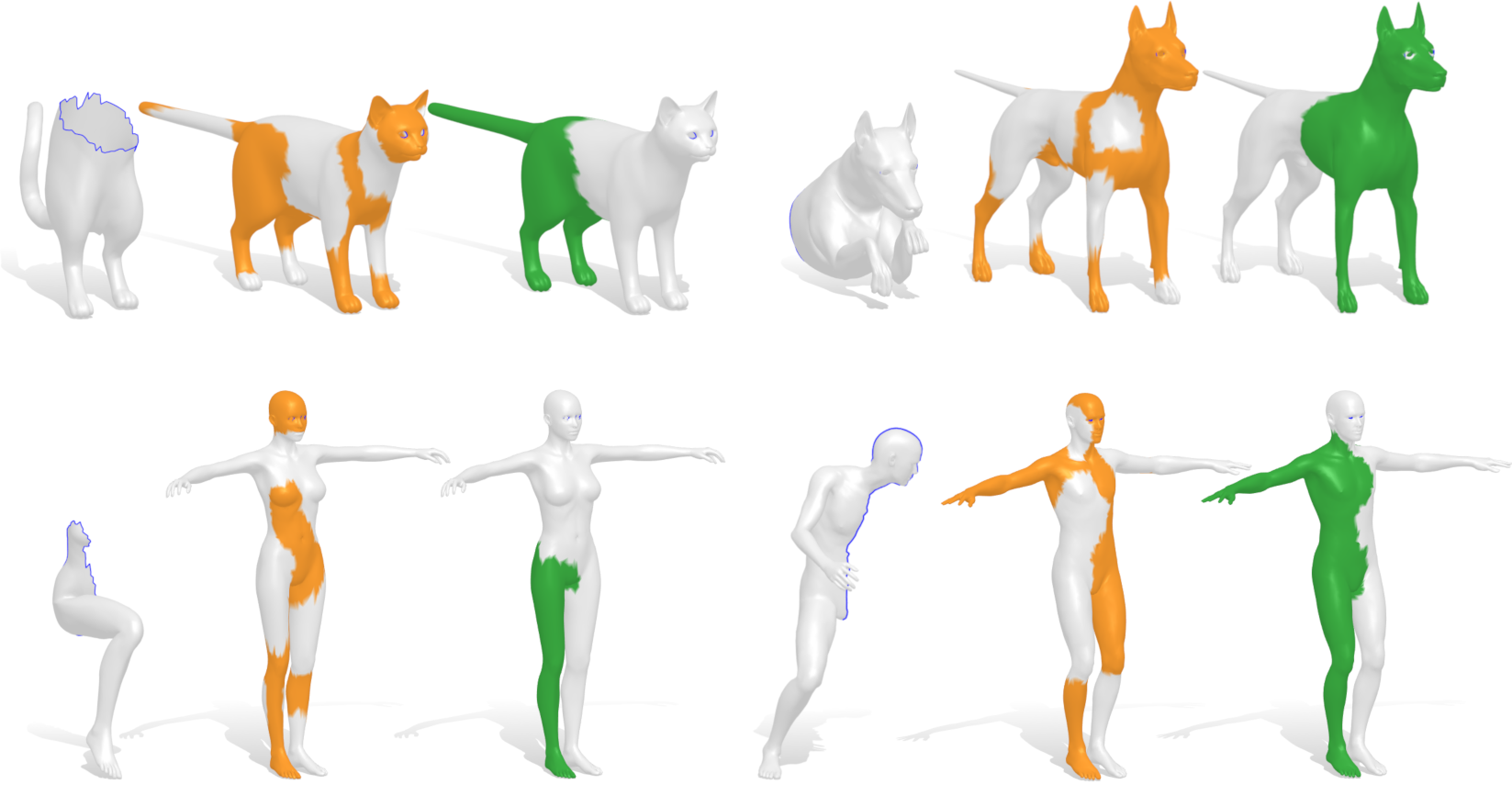}
\put(17.3,31.3){\footnotesize 0.38} 
\put(68.5,31.3){\footnotesize 0.62} 
\put(17.3,0.5){\footnotesize 0.39} 
\put(68.5,0.5){\footnotesize 0.48} 
\put(37.1,31.3){\footnotesize \textbf{0.96}} 
\put(84.9,31.3){\footnotesize \textbf{0.96}} 
\put(37.1,0.5){\footnotesize \textbf{0.79}} 
\put(84.9,0.5){\footnotesize \textbf{0.77}} 
\end{overpic}
  \caption{\label{fig:pfm}Comparisons with PFM on the SHREC'16 benchmark. For each example we show the partial query, the PFM solution (orange), and our solution (green). The numerical score below each solution denotes the intersection-over-union with respect to the ground truth mask. Note how PFM tends to mix-up the bilateral symmetries of the shapes (left-right and front-back), since it relies on local descriptors that can not discriminate orientation.}
\end{figure}

For PFM we used public code released by the authors with default parameters. As a baseline, we also evaluate the co-segmentation approach of~\cite{toldo2009bag} with SHOT descriptors \cite{tombari2010unique}. Quantitative and qualitative comparisons are reported in Figure~\ref{fig:iou} and Figure~\ref{fig:pfm} respectively.

%
 
\begin{figure}[b]
  \centering
\begin{overpic}
[trim=0cm 0cm 0cm 0cm,clip,width=1.0\linewidth]{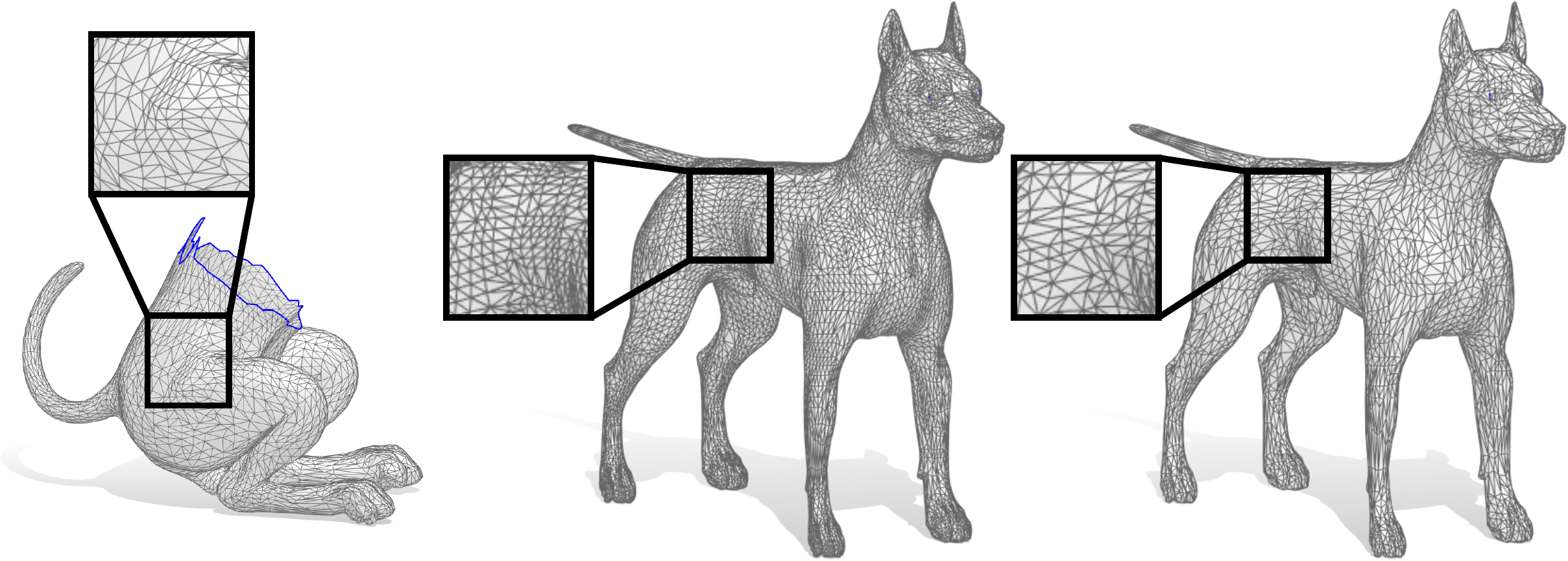}
\put(40,29){\footnotesize $n=10K$}
\put(77.5,29){\footnotesize $n=5K$}
\end{overpic}
%
%
\definecolor{mycolor1}{rgb}{0.00000,0.44700,0.74100}%
\definecolor{mycolor2}{rgb}{0.85000,0.32500,0.09800}%
\definecolor{mycolor3}{rgb}{0.92900,0.69400,0.12500}%
\definecolor{mycolor4}{rgb}{0.49400,0.18400,0.55600}%
\definecolor{mycolor5}{rgb}{0.2,0.5882,0.2235}%
\definecolor{mycolor6}{rgb}{0.30100,0.74500,0.93300}%
\definecolor{mycolor7}{rgb}{0.63500,0.07800,0.18400}%
\definecolor{mycolor8}{rgb}{0.9216,0.5765,0.1725}%
\pgfplotsset{ 
compat=1.11,
legend image code/.code={
\draw[mark repeat=2,mark phase=2]
plot coordinates {
(0cm,0cm)
(0.15cm,0cm)        
(0.3cm,0cm)         
};%
}
}
\begin{tikzpicture}

\begin{axis}[%
width=0.2\linewidth,
height=0.25\linewidth,
scale only axis,
every x tick label/.append style={font=\color{black}, font=\tiny},
every y tick label/.append style={font=\color{black}, font=\tiny},
xmin=0.4,
xmax=1,
xlabel style={font=\color{white!15!black}},
xlabel={\footnotesize IoU},
xlabel style={at={(axis description cs:0.5,-0.04)},anchor=north},
ylabel style={at={(axis description cs:-0.28,0.5)},anchor=north},
ymin=0,
ymax=1,
ytick={0,0.2,0.4,0.6,0.8,1},
yticklabels={0,,,,,1},
xtick={0,0.2,0.4,0.6,0.8,1},
xticklabels={0,,0.4,,,1},
ymajorgrids,
xmajorgrids,
axis background/.style={fill=white},
legend style={legend cell align=left, align=left, draw=white!15!black, text height=0.9ex, fill=white, fill opacity=0.6, draw opacity=1, text opacity=1, at={(0.68,0.01)}, anchor=south east},
legend style={inner sep=0pt}
]


 \addplot [color=mycolor8, dotted, line width=1.2pt]
   table[row sep=crcr]{%
 0.1	0.991666666666667\\
 0.25	0.991666666666667\\
 0.3	0.983333333333333\\
 0.35	0.966666666666667\\
 0.4	0.966666666666667\\
 0.45	0.958333333333333\\
 0.5	0.9\\
 0.55	0.875\\
 0.6	0.833333333333333\\
 0.65	0.8\\
 0.7	0.741666666666667\\
 0.75	0.65\\
 0.82	0.533333333333333\\
 0.84	0.533333333333333\\
 0.86	0.516666666666667\\
 0.88	0.508333333333333\\
 0.9	0.45\\
 0.91	0.45\\
 0.92	0.433333333333333\\
 0.94	0.383333333333333\\
 0.95	0.35\\
 0.96	0.325\\
 0.97	0.283333333333333\\
 0.98	0.233333333333333\\
 0.99	0.0666666666666667\\
 1	0\\
 };
 \addlegendentry{\footnotesize PFM \cite{rodola2017partial}}

\addplot [color=mycolor8, line width=1.2pt]
  table[row sep=crcr]{%
0.1	0.991666666666667\\
0.15	0.983333333333333\\
0.2	0.983333333333333\\
0.25	0.975\\
0.3	0.95\\
0.35	0.933333333333333\\
0.4	0.9\\
0.45	0.875\\
0.5	0.808333333333333\\
0.6	0.691666666666667\\
0.65	0.616666666666667\\
0.7	0.516666666666667\\
0.75	0.433333333333333\\
0.8	0.375\\
0.82	0.366666666666667\\
0.84	0.341666666666667\\
0.86	0.308333333333333\\
0.88	0.3\\
0.9	0.275\\
0.91	0.266666666666667\\
0.94	0.191666666666667\\
0.96	0.158333333333333\\
0.97	0.116666666666667\\
0.98	0.1\\
0.99	0.0249999999999999\\
1	0\\
};
\addlegendentry{\footnotesize PFM (remesh)}

\addplot [color=mycolor5, dotted, line width=2.0pt]
   table[row sep=crcr]{%
 0.1	1\\
 0.15	0.975\\
 0.2	0.958333333333333\\
 0.25	0.95\\
 0.3	0.916666666666667\\
 0.35	0.9\\
 0.4	0.875\\
 0.45	0.875\\
 0.6	0.725\\
 0.65	0.691666666666667\\
 0.75	0.658333333333333\\
 0.8	0.608333333333333\\
 0.82	0.608333333333333\\
 0.84	0.575\\
 0.86	0.525\\
 0.88	0.508333333333333\\
 0.9	0.483333333333333\\
 0.91	0.475\\
 0.92	0.475\\
 0.93	0.441666666666667\\
 0.94	0.425\\
 0.95	0.4\\
 0.96	0.333333333333333\\
 0.97	0.241666666666667\\
 0.99	0.0249999999999999\\
 1	0\\
 };
 \addlegendentry{\footnotesize \textbf{Ours}}

\addplot [color=mycolor5, line width=2.0pt]
  table[row sep=crcr]{%
0.1	0.991666666666667\\
0.15	0.983333333333333\\
0.25	0.933333333333333\\
0.3	0.891666666666667\\
0.35	0.858333333333333\\
0.4	0.85\\
0.45	0.825\\
0.5	0.791666666666667\\
0.55	0.75\\
0.6	0.733333333333333\\
0.65	0.691666666666667\\
0.75	0.675\\
0.8	0.608333333333333\\
0.82	0.583333333333333\\
0.84	0.566666666666667\\
0.86	0.508333333333333\\
0.88	0.491666666666667\\
0.9	0.466666666666667\\
0.91	0.466666666666667\\
0.92	0.45\\
0.94	0.383333333333333\\
0.95	0.366666666666667\\
0.96	0.358333333333333\\
0.97	0.241666666666667\\
0.98	0.15\\
0.99	0.0416666666666667\\
1	0\\
};
\addlegendentry{\footnotesize \textbf{Ours} (remesh)}
\legend{}
\end{axis}
\end{tikzpicture}%
\begin{overpic}
[trim=0cm -1.5cm 0cm 0cm,clip,width=0.73\linewidth]{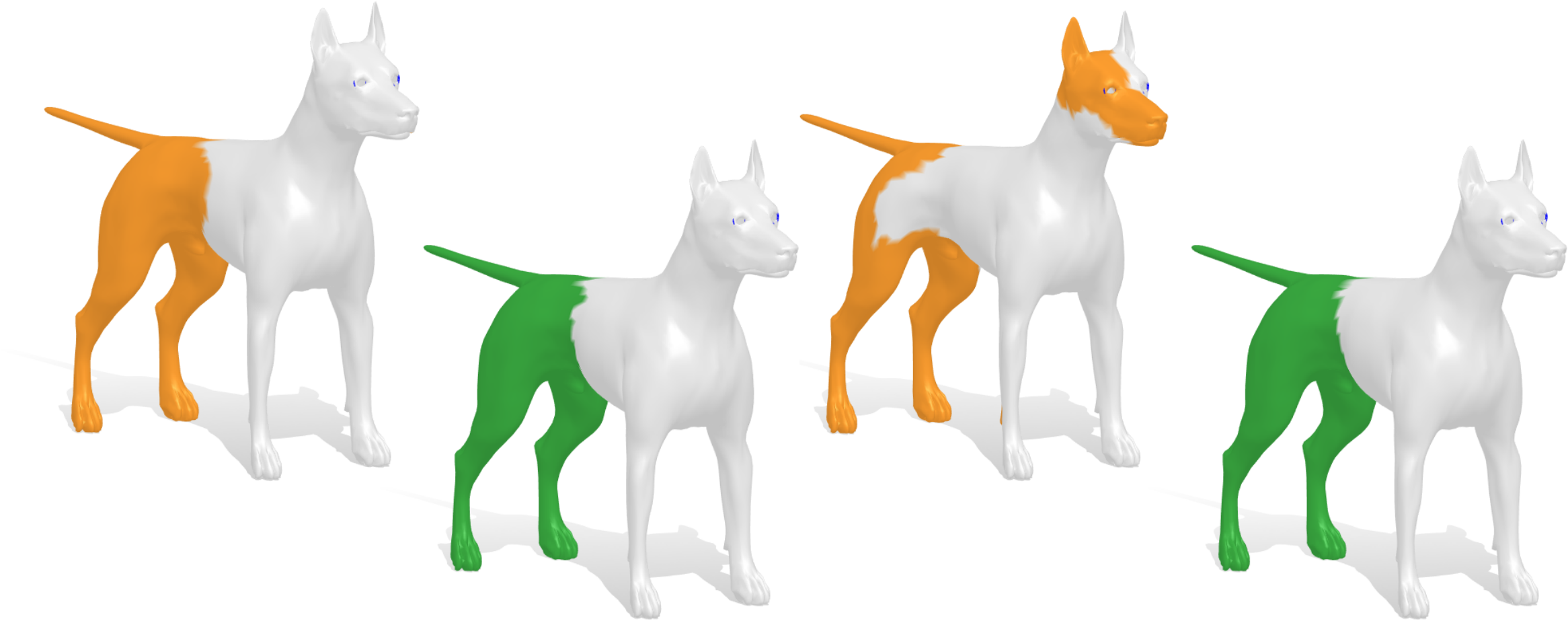}
\put(05,1.7){\footnotesize PFM \cite{rodola2017partial}}
\put(35,1.7){\footnotesize \textbf{Ours}}
\put(54.5,1.7){\footnotesize PFM \cite{rodola2017partial}}
\put(84.5,1.7){\footnotesize \textbf{Ours}}
\end{overpic}
  \caption{\label{fig:sampling}Robustness to changes in vertex density. We show the results of PFM (in orange) and our method (in green) over two versions of the entire SHREC'16 dataset: a version where all meshes have similar resolution (dotted curves in the plots), and one where mild remeshing has been applied (solid curves). PFM shows a performance drop of 20\%, while our method remains stable.}
\end{figure}
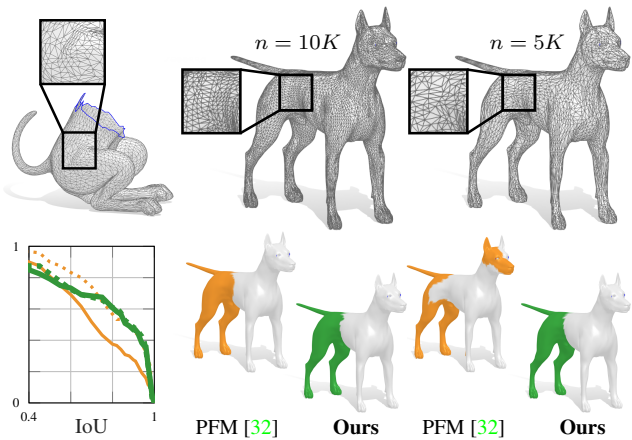

\vspace{1ex}\noindent\textbf{Robustness to sampling.}
Our method exhibits robustness to changes in resolution (see Figure~\ref{fig:sampling} for a full evaluation). We attribute this to the fact that eigenvalues are integral quantities (see Eq.~\ref{eq:de}). By contrast, PFM fully relies upon a data term based on local point descriptors, which are extremely sensitive to local meshing.


\vspace{1ex}\noindent\textbf{Multiple pieces.}
We highlight another useful property derived from working with potentials. In Figure~\ref{fig:puzzles} we show examples where the input 3D query consists of {\em multiple} disconnected parts (the ``non-rigid puzzle'' setting with non-overlapping pieces \cite{litany2016non}). By taking their union and treating them as one single shape $\Y$ in our optimization problem, we are still able to correctly identify the distinct portions on the full shape without any special adjustment to our algorithm.

\vspace{1ex}\noindent\textbf{Implementation details.}
In all our tests we use $k=20$ eigenvalues.
As initialization for $\v$, we use Gaussians centered around $m=20$ Euclidean farthest samples on $\X$, with variance equal to $\sqrt{\mathrm{Area}(\X)}$ and $2\sqrt{\mathrm{Area}(\X)}$, leading to $2m$ different initializations which we optimize in parallel. To these, we also add the solutions produced by the baseline.
%
To choose the final solution we rely on Property~\ref{thm:align}. For each found $v$, we compute projections of SHOT descriptors $F: \X \to \mathbb{R}$ 
 over the first $5$ Hamiltonian eigenfunctions (squared), $c_{i} = \int_\X \psi_i^2(x) F(x) \mathrm{d}x$ for $i=1,\dots,5$. We then compute coefficients $d_i$ on $\Y$ similarly by using the Dirichlet eigenfunctions, and keep the $v$ which yields minimum distance $\sum_{i=1}^5 (d_i-c_i)^2$. By Property~\ref{thm:align}, we expect these coefficients to be the same for the correct solution.


\subsection{Isospectralization}\label{sec:isosp}
As a second application we address the shape-from-spectrum problem. This task was recently introduced in \cite{isosp} and is phrased as follows: Given as input a short sequence of Laplacian eigenvalues $\{\mu_i\}_{i=1}^{k}$ of an {\em unknown} shape $\Y$, recover a geometric embedding of $\Y$. 
We stress that the Laplacian $\Delta_\Y$ is {\em not} given; if given, it would lead to a shape-from-operator problem \cite{boscaini2015shape}. Instead, the goal here is to recover the shape directly from the eigenvalues alone.

We model this problem by minimizing Eq.~\eqref{eq:prob2} with the following input. As full domain $\X$ we consider a finite portion of the plane $\X\subset\mathbb{R}^2$ of size $n\times n$, sampled on a Cartesian grid. The input eigenvalues $\{\mu_i\}$ are assumed to be coming from an unknown region $\R\subset\X$, which we identify by the potential $v_\tau$ on $\X$. Therefore, our optimization variables are simply $\v\in(0,\tau)^{n\times n}$ -- essentially, an image.

\begin{property}
We can recover shapes with \textbf{unknown topology}, \ie, no restricting assumptions are made on the topology of the sought shape.
\label{thm:topo}
\end{property}

For example, if the shape to be recovered has holes, we do not need to know this fact beforehand to properly initialize the optimization.

\vspace{1ex}\noindent\textbf{Comparisons.}
In Figure~\ref{fig:isosp} we compare directly with the isospectralization technique of Cosmo~\etal~\cite{isosp}; to our knowledge, this is the only existing approach attempting to solve shape reconstruction purely from eigenvalues. Our method is different in many respects. 

The most crucial difference lies in the fact that in \cite{isosp} the authors optimize for a {\em deformation field} to apply to an initial template shape, which is assumed to be given as an input. In contrast, with Problem~\eqref{eq:prob2} we simply minimize over a vector space of scalar functions.
The initial shape used in \cite{isosp} is also required to have the correct topology (\eg, disc-like or annulus-like), while by Property~\ref{thm:topo} our optimization is completely oblivious to the topology of the shape to recover. We illustrate this with an example in the rightmost column of Figure~\ref{fig:isosp}, where we directly compare with \cite{isosp}.


Finally, optimizing over embedding coordinates as in \cite{isosp} requires the adoption of additional regularizers to avoid flipped triangles, and careful parameter tuning to avoid rough shapes and collapsing edges; this makes their energy more difficult to minimize, which is done using stochastic optimization tools \cite{kingma2014adam} and automatic differentiation as implemented in TensorFlow~\cite{tf}. By optimizing over scalar functions instead of deforming a given mesh, we do not encounter any of these issues; for this reason, our optimization can be made much simpler as detailed in Section~\ref{sec:method}.

\begin{figure}[t]
  \centering
\begin{overpic}
[trim=0cm 0cm 0cm 0cm,clip,width=1.0\linewidth]{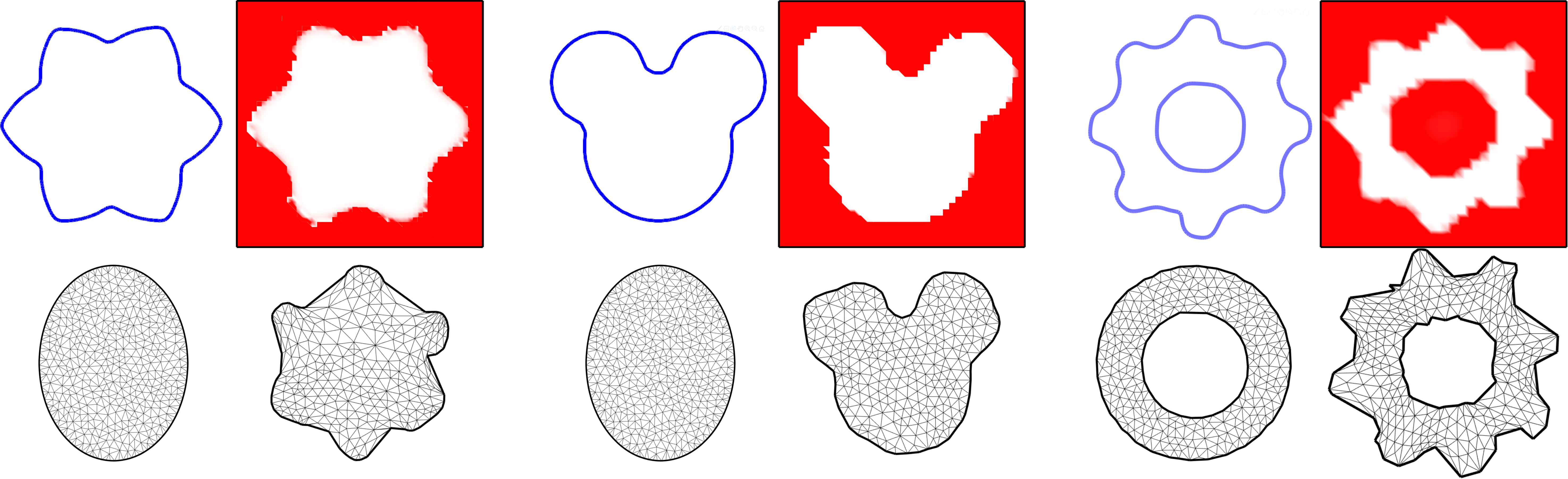}
\end{overpic}
  \caption{\label{fig:isosp}Shape recovery comparisons with \cite{isosp}. We show the ground truth shapes (blue outline) and our solutions (potential on the square, growing from white to red), all obtained in 2-3 minutes from $k=50$ input eigenvalues. We always use a constant initialization. On the bottom we show results for \cite{isosp} with their initial templates, which are required to always have the correct topology.}
\end{figure}
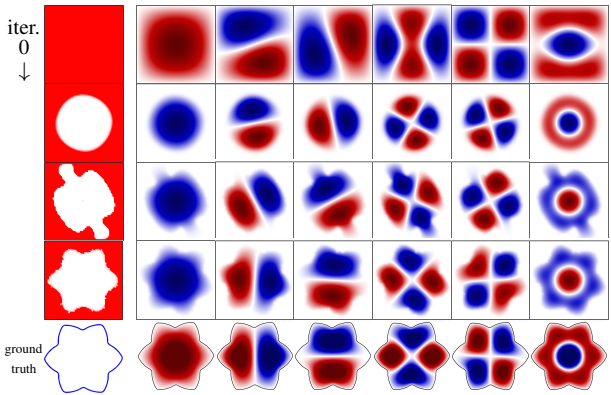
\begin{figure}[t]
  \centering
\begin{overpic}
[trim=0cm 0cm 0cm 0cm,clip,width=0.9\linewidth]{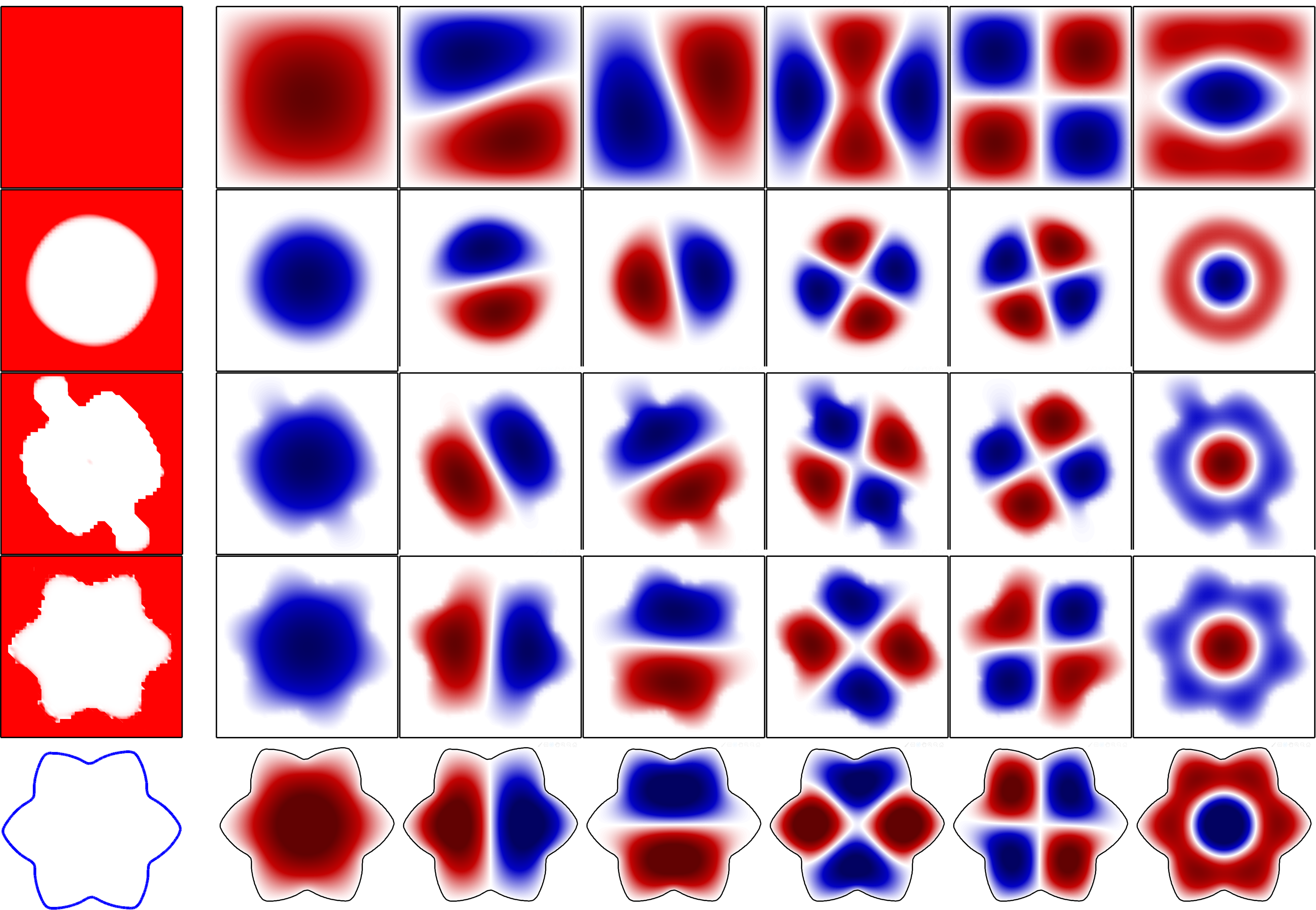}
\put(-6.5,63.5){\footnotesize iter.}
\put(-4.5,60.0){\footnotesize 0}
\put(-4.5,56.0){\footnotesize $\downarrow$}
\put(-7,7.0){\tiny ground}
\put(-5.8,3.5){\tiny truth}
\end{overpic}
  \caption{\label{fig:star}Shape recovery across iterations. Starting from a sequence of just $k=5$ Laplacian eigenvalues, our method recovers the correct shape in less than $1$ minute. The first column shows our solution, evolving from top to bottom. On the right we also show how the eigenfunctions (one per column) change across the iterations; observe how at convergence, they align with those of the ground truth shape (up to rotation).}
\end{figure}

\vspace{1ex}\noindent\textbf{Implementation details.}
The planar domain $\X$ is discretized as a $30 \times 30$ grid with uniform triangles.
We initialize simply with the constant potential $\v = \frac{\tau}{4} \mathbf{1}$, in order to emphasize the ability of our method to recover shape entirely from scratch irrespective of topology. We put this in contrast with \cite{isosp}, which needs an initial template as initialization, as mentioned above. See Figure~\ref{fig:star} for a visualization across some iterations of our optimization process.

\section{Conclusions}
\vspace{-0.5ex}
In this paper we presented a new approach for partial shape similarity, based on aligning short sequences of eigenvalues computed on the given shapes. Our approach is correspondence-free, is invariant to non-rigid transformations, it can be solved efficiently, and only makes use of descriptors for post-processing refinement. Our formulation is also general enough to be applied to other relevant tasks in vision and graphics. We demonstrated this on a challenging shape-from-spectrum recovery problem, where it compares favorably with respect to a very recent method.

\begin{figure}[bt]
  \centering
\begin{overpic}
[trim=0cm 0cm 0cm 0cm,clip,width=0.85\linewidth]{./figures/fail.png}
\put(0,16.5){\footnotesize 0.51}
\put(51,16.5){\footnotesize 0.13}
\end{overpic}
\begin{overpic}
[trim=0cm 0cm 0cm -2.0cm,clip,width=0.85\linewidth]{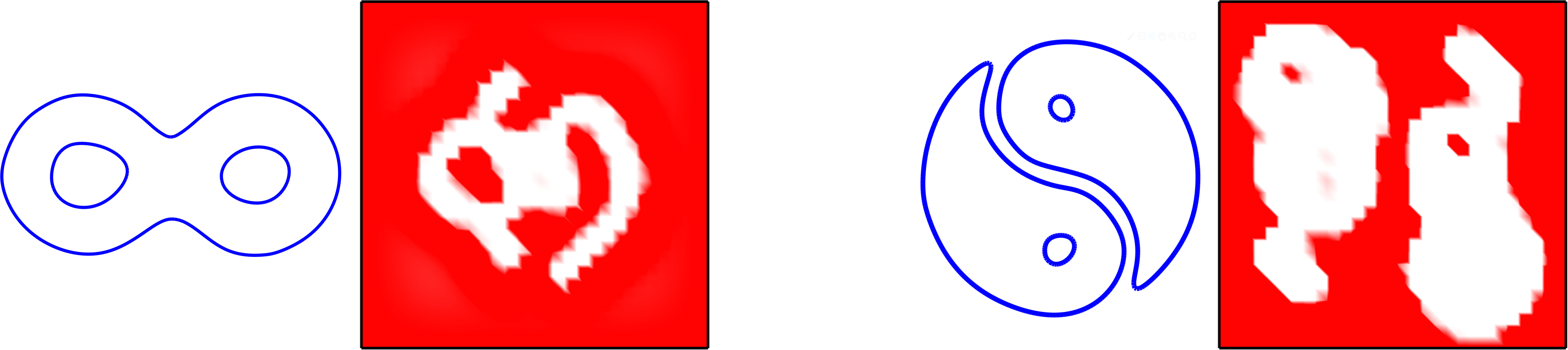}
\end{overpic}
  \caption{\label{fig:fail}Failure cases. {\em Top}: On the left example, the eigenvalues did not align well at the local optimum; as a result, the partial shape was localized only roughly. On the right, the found region is wrongly classified as correct since it has similar eigenvalues to the partial input within the limited bandwidth. In both cases we report the IoU. {\em Bottom}: Shape-from-spectrum recovery of shapes with non-trivial topology can be difficult to achieve, due to the narrow bandwidth and first-order discretization of the operators.}
\end{figure}

\vspace{1ex}\noindent\textbf{Limitations and future work.}
We did observe some failure cases during our tests (see Figure~\ref{fig:fail}). In the partial similarity setting, this typically happens whenever the chosen bandwidth (\ie, the number $k$ of input eigenvalues) is too small for discriminating regions that only differ at medium-high frequencies; \eg, a human arm might be confused with a leg. This can be remedied to some extent by increasing the value of $k$. For large values ($k>100$), however, a more accurate discretization of the operator $\Delta$ should be used in place of the cotangent formulas of Eq.~\eqref{eq:cotan}, to avoid artifacts related to mesh tessellation. Using second- or third-order FEM discretization \cite{ciarlet2002finite} might be a promising solution. 

Furthermore, while we only showed 2D examples of shape-from-spectrum recovery, the exact same algorithm can be used invariantly for reconstructing the geometry of 3D (volumetric) shapes. However, this would require solving for a potential defined in $\mathbb{R}^3$, for which efficient data structures (such as octrees) could be necessary. We leave this exciting possibility to future work.

Finally, a theoretical question that we left open concerns the empirical observation that eigenfunction alignment is induced in many practical situations. We believe that this requires a deeper understanding, with potentially profound consequences in many applied domains.

{\footnotesize
\section*{Acknowledgments} 
\vspace{-1.5ex}
\noindent
The authors wish to thank Luca Cosmo, Luca Balsanelli and Jing Ren for useful discussions and the technical help. AR and ER are supported by the ERC Starting Grant no. 802554 (SPECGEO). Parts of this work were supported by the KAUST OSR Award No. CRG-2017-3426, a gift from the NVIDIA Corporation and the ERC Starting Grant No. 758800 (EXPROTEA).
}

{\small
\bibliographystyle{ieee}
\bibliography{egbib}
}

\end{document}